\documentclass[%
 twocolumn,
 amsmath,amssymb,
 aps,
 pra,
floatfix,
]{revtex4-2}
\usepackage[utf8]{inputenc}
\usepackage{amsmath, dsfont}
\usepackage{dcolumn}
\usepackage{bm}
\usepackage{empheq}
\usepackage{float}
\usepackage{tikz}
\usepackage{graphicx} 
\graphicspath{ {./figs/} }
\usepackage{mathtools}
\usepackage{amsfonts}
\usepackage{amssymb}
\usepackage[margin=1.0in]{geometry}
\usepackage{dcolumn}
\usetikzlibrary{patterns}
\newfont{\sfd}{cmssdc10 scaled\magstep0}
\newfont{\cmu}{cmu10 scaled\magstep0}

\def\newpic#1{}

\newcommand{\opOne}{\mathds{1}}

\newcommand{\imag}[1]{\mbox{Im}[#1]}

\def\rv_#1{| #1 \rangle}
\def\lv_#1{\langle #1 |}
\def\av_#1^#2{\langle #1 | #2 | #1 \rangle}
\def\ov_#1^#2{\langle #1 | #2 \rangle}
\def\avop#1{\langle #1 \rangle}

\def\psirt_#1{\psi_{#1}({\bf r},t)}
\def\phirt_#1{\varphi_{#1}({\bf r},t)}
\def\Psirt_#1{\Psi_{#1}({\bf r},t)}
\def\Phirt_#1{\Phi_{#1}({\bf r},t)}
\def\psirtstar_#1{\psi^\ast_{#1}({\bf r},t)}
\def\phirtstar_#1{\varphi^\ast_{#1}({\bf r},t)}
\def\Psirtstar_#1{\Psi^\ast_{#1}({\bf r},t)}
\def\Phirtstar_#1{\Phi^\ast_{#1}({\bf r},t)}
\def\psir_#1{\psi_{#1}({\bf r})}
\def\phir_#1{\varphi_{#1}({\bf r})}
\def\Psir_#1{\Psi_{#1}({\bf r})}
\def\Phir_#1{\Phi_{#1}({\bf r})}
\def\psirstar_#1{\psi^\ast_{#1}({\bf r})}
\def\phirstar_#1{\varphi^\ast_{#1}({\bf r})}
\def\Psirstar_#1{\Psistar_{#1}({\bf r})}
\def\Phirstar_#1{\Phistar_{#1}({\bf r})}

\def\I{{\bf I}}

\def\L{{\bf L}}

\def\T{{\bf T}}
\def\T{{\bf T}}

\def\opd{\hat{d}}

\def\opH{\hat{H}}

\def\opL{\hat{L}}

\def\opV{\hat{V}}

\def\oprho{\hat{\rho}}

\def\calD{\mathcal{D}}
\def\calE{\mathcal{E}}

\def\calS{\mathcal{S}}

\def\calK{\mathcal{K}}

\def\calI{\mathcal{I}}

\def\lrarrow{~\lower.2ex\hbox{$\rightarrow$}\kern-2.4ex\raise.7ex\hbox{$\leftarrow$}~}
\def\rlarrow{~\lower.2ex\hbox{$\leftarrow$}\kern-2.3ex\raise.7ex\hbox{$\rightarrow$}~}

\def\caltE{\tilde{\mathcal{E}}}

\usepackage[export]{adjustbox}
\usepackage[nameinlink,noabbrev,capitalize]{cleveref}
\crefformat{equation}{~(#2#1#3)}
\creflabelformat{section}{~(#2#1#3)}

\newcommand{\dprime}{{\prime\prime}}

\begin{document}
\title{Perturbative theory of ensemble-averaged atomic dynamics in fluctuating laser fields}
\author{Tejaswi Katravulapally and L. A.A. Nikolopoulos}
 \email{Correspondence: Lampros.Nikolopoulos@dcu.ie}
\affiliation{%
School of Physical Sciences, Dublin City University, Dublin 9, Ireland.\\
}%
\date{\today}
\begin{abstract}
We have developed a perturbative method to model the resonant ionization of atomic systems in fluctuating laser fields. The perturbative method is based on an expansion in terms of the multitime cumulants, a suitable combination of moments (field's coherence functions), used to represent the field's statistical properties. The second-order truncated expansion is expressed in terms of the radiation's power spectrum and the intensity autocorrelation function. We investigate the range of validity of the model in terms of the field's coherence temporal length and peak intensity and have compared the results with conventional Monte-Carlo calculations. We apply the theory in the case of a near-resonant ionization of the Helium $2s2p$ autoionizing state with a SASE FEL pulse with square-exponentially dependent 1st-order coherence function. The ionization lineshape profile acquires a Voight profile; the degree of the Gaussian or Lorentzian character of which to depend crucially on the field's coherence time. 
\begin{description}
\item[Keywords:] FEL, Autoionization, Resonance ionization, stochastic differential equations, Monte Carlo.
\item[Usage]
\item[PACS numbers]
\item[Structure]
\end{description}
\end{abstract}


\maketitle


\section{Introduction}
\noindent
Following the advent of the Free-Electron Laser (FEL) sources the dynamics of atomic systems under a randomly varying short-wavelength radiation is of research interest area for about more than a decade now \cite{wabnitz:2002,ackermann:2007,young:2010}.
From a mathematical standpoint the field's randomness renders the system's equations-of-motions (EOMs) to a \textit{stochastic differential equations} problem with the dynamical quantities treated as stochastic processes \cite{vankampen:1976,papoulis:2002,lax:2006}. The practical problem faced is to describe the statistics of these dynamical quantities (excited population, ionization yield, electron  kinetic/angular spectrum, absorption spectrum etc...) given the radiation's statistical properties. This typical probabilistic problem was dealt theoretically in detail in a very alike context soon after the first laser sources of long-wavelength radiation appeared and experimental data became available. The interested reader can find rigorous studies of these kind of problems in the early works in the field and references therein ~\cite{agarwal:1970,georges:1979,zoller:1982,eberly:1984}. With the later development of lasers with well stabilized amplitude and phase (e.g. Ti:Sapphire) the weakened fluctuations proved to play negligible role in the excitation/ionization dynamics making a statistical description redudant. Nowadays, a quite similar scenario emerged for the FEL's radiation which exhibits strong temporal fluctuations, thus naturally triggering a revived interest about the effects of randomness in atomic photoionization dynamics \cite{santra:2008,young:2010,nikolopoulos:2011,nikolopoulos:2012,lambropoulos:2012,Mouloudakis}. A class of experimental schemes (multi-shot) measure data which have been generated over many pulses which differ each other in a random manner; therefore, inevitably, these measurements provide the \textit{averaged} values of the experimental observables. In mathematical parlance, they represent an \textit{ensemble} average over the field's fluctuations. This simple fact brings again at the forefront a statistical description of the atomic dynamics.  
 
Our purpose is to present a systematic perturbation theory of near-resonant photoionization processes which takes into account the fluctuation statistics of the FEL radiation 
[Fig.~(\ref{fig:fig1})]. We assume the FEL pulses as described in the work by Krinsky and Li \cite{Krinsky1}, which model the self-amplified spontaneous emmision (SASE) start-up process as a shot-noise stochastic process \cite{Rice1944,saldin_book:2000,kim:2017}. According to this model, the first-order coherence function of the FEL radiation possess a square-exponential (Gaussian) 
time dependence, $\sim e^{-(t_1-t_2)^2/2\tau_c^2}$ (this is distinctly different from the usually assumed, $\sim e^{-|t_1-t_2|/2\tau_c}$, of the early lasers of longer-wavelength). The Gaussian-like dependence has been treated to date only within a Monte-Carlo (MC) algorithm type of calculations 
\cite{santra:2008,lambropoulos:2012}. One of the outcomes of the present formulation is the following: as usual the infinite term pertubation series is exact, but, unless some simplified cases, one is led to truncate the expansion. The perturbative method clearly sets the range of validity of the truncated expansion beyond any doubt. A second outcome is that the lowest-order term of the truncated expansion is expressed in terms of quantitities directly relevant to experimental observables, namely its temporal mean intensity and its power spectrum; more accurately, the autocorrelation (AC) functions of the time-integrated electric field amplitude and intensity, $\avop{\calE(t)\calE^*(t^\prime)}$ and 
$\avop{\calI(t)\calI(t^\prime)}$, respectively. The intensity coherence function is essential to be included; for the reason that different random fields may share the same 1st-AC function but not the second. This dependance of the dynamics on these AC functions becomes stronger for FEL fields obeying Gaussian statistics  and especially at conditions where stationarity of the averages may be assumed (long pulses relative the field's coherence time) \cite{goodman:1985}.

The structure of the presentation is as follows: in Sec.~II we start with a brief discourse on 
the EOMs theory in stochastic fields followed from the detailed development of the particular perturbative method. In this section we arrive at an exact differential deterministic equation for the ensemble-averaged density matrix driven by a time-dependent perturbation expansion. The ionization scheme assumed is given in Fig.~(\ref{fig:fig1}). Apart from a 'short' correlation-time requirement not any other particular assumption for the field is made. We also evaluate the first two non-vanishing terms of the series expansion and obtain the Eqs.~(\ref{eq:avg-b}) and (\ref{eq:avg-c}) for the bound and the singly-ionized states of the system. 
These equations are the main result of this work. In Sec.~III the terms of the truncated expansion are specialized for fields obeying Gaussian statistics for the cases where the first-order AC function is square-exponentially and exponentially correlated; the former corresponds to the FEL radiation model mentioned earlier whereas the latter is more suitable for various models of long-wavelength laser radiation \cite{wilhelmi:1992,avan:1977}. In Sec.~IV we apply the method in the near-resonant ionization regime via the Helium $2s2p$ autoionization state (~$\sim 60.2$ eV) for two FEL pulses, of different intensity FWHM duration 
$\sim 11.7$ and $75$ fs, and compare the results with a MC method. 

 In the Conclusions section we discuss limitations of the present method and further work along similar lines. Finally in the appendix we have relegated some of the mathematical formulas outside the main text. 

For the presentation of the formulas, atomic units are used throughout the text; For the reader's convenience other units (e.g. eV, fs) are used in the discussion section and the figures caption.  

The less mathematically inclined reader may skip Sec. II all together and focus on the more applied part of the text. The reader who is familiar with the theory of density-matrix resonant ionization and is mostly interested in the averaging method should focus to Sec. IIb.

\section{Formulation}

In the below we develop a perturbation method for the averaged density matrix equations which describes the excitation and ionization of an atomic system via an autoionizing state under a randomly fluctuating field. The discussion is kept general for any fluctuating radiation field, apart from the main assumption that the field's coherence time should be shorter than any other characteristic times relevant to the systems dynamics; let's denote this coherence time as $\tau_c$.

The first step to solving the stochastic EOMs is to choose the statistical model for the (external) radiation field. Regarded as a stochastic processes it cannot be modelled via analytical functions, as is customarily the case for a fluctuation-free laser field. 
It maybe represented either by its multitime joint-probability distribution function for the electric field, or equivalently via its multi-time average moments (coherences). Mathematically the field may be treated either as a real variable or more conveniently via its complex representation \cite{mandelwolf:1995}. For the appropriate modelling the standard laser theory for the field may be used or available experimental data (average power spectrum, intensity AC spectrogram) \cite{kim:2017,Pfeifer2010}. In either case one is of possesion of a number of random time-series of the field, $E^{(j)}(t_i), i =1,2,\cdots ; j=1,2,\cdots,$ with the required statistical properties met. 

Next, a suitable method should be chosen to solve the EOMs. Avoiding to delve into the variety of methods available we briefly refer only on the brute-force MC route where one uses the random time series of the electric field, $E^{(j)}(t_i)$ as is. For each of these realizations either the appropriate time-dependent Schr\"{o}dinger equation (TDSE) or the density matrix equation (TDDM), is solved using the standard calculational methods employed for deterministic laser fields. Gathering the results of these calculations, we can obtain the average of the desired observable (e.g. ionization yield, electron kinetic/angular spectrum, etc..). This approach, although it is benefited from the use of the accumulated experience and methods, it has the downside that it must be repeated many times; to reach a trusted result the total number of realizations might be of the order of hundreds or thousands which for the more demanding problems might impractical.    

The alternative method, followed here, is to achieve the same goal by performing a statistical average of the stochastic equations themselves; the hope is that a sure set of EOMs can be obtained for the statistical average of the observable under question. The immediate benefit of the averaging process is a significant computational gain to calculate the desired averages; a second benefit is that although the averaged equations have a similar structure with their stochastic counterpart a clearer 
insight of the essential quantities participating in the dynamics might be obtained. These two properties have been our motivation towards developing an averaging set of EOMs for near-resonant atomic interactions with FEL fields. 


Turning now to more specific aspects of the averaging method, we need to clarify some terminology to be used later in the text. As \textit{statistical (ensemble) average} over the variations of a random quantity the following is meant: assuming a quantity, $f(x)$, dependent on the random process $x(t)$ then its statistical multitime average is defined by $\avop{f(t_1,t_2,\cdots,t_n)}\equiv \int dx_1dx_2 \cdots dx_n p_n(x_1,x_2,\cdots,x_n) f(x_1,x_2\cdots,x_n)$, where $x_i=x(t_n)$ and $p_n(x_1,x_2,\cdots x_n)$ is the $n$-th order joint \textit{probability density distribution} of the $x(t)$ process. Very frequently the physical conditions allows the description in terms of an important class of stochastic processes which require the knowledge of only the first and second moments, $\avop{x(t_1)}$ and $\avop{x(t_1)x(t_2)}$, known as the mean and the autocorrelation functions, respectively (for radiation field these are known as statistical 1st- and 2nd- coherence functions). In such cases we say that the process obeys, \textit{Gaussian} statistics. Generally, for a random process the AC function, $\avop{x(t_1)x(t_2)}$, decays in the course of the time; its rate of decay largely sets the autocorrelation time, $\tau_c$; its value for $t_1=t_2$ is directly related with the field's \textit{average power} (actually is the Fourier integral of the AC function). When the mean value is constant and the AC function has a dependence of the form $|t_1-t_2|$ the random process is (2nd-order) \textit{stationary}; in relation to this, strictly speaking, a fluctuating pulse maybe represented as a stationary random processes only asymptotically in the limit of infinitely long pulses; nevertheless, in practice a field of a large number of cycles may be approximately assumed as stationary and can be represented by a well defined time-independent power spectum.    

Another point to be clarified is that a proper formulation for the statistical average of quantum mechanical quantities requires a density matrix representation of the system's states (the field is treated classically); this is due to the quantum postulates for the expectation values of an observable; A Hilbert representation of the system's state expresses the population and coherences as amplitude products, say $\sim c_n(t)c^*_{n^\prime}(t)$. However, generally, for an ensemble average $\avop{c_{n}(t)c^*_{n'}(t)} \neq \avop{c_{n}(t)} \avop{c^*_{n'}(t)}$; so, since only $\avop{c_{n}(t)}$ can be calculated from an averaged TDSE formulation inevitably such approach would lead to erroneous results. 

The above argument calls for the ensemble-averaged matrix elements $\rho(t)$ governed by the Liouville equation \cite{Haar}, $\imath \dot{\rho}(t) = [\opH_a + \opV(t), \rho(t)]$, where $\opH_a$ is the time-independent field-free atomic Hamiltonian operator and $\opV(t)$ representing the random external potential. Assuming the ensemble average over the external random fluctuations we have,
\begin{equation}
\imath \avop{\dot{\rho}(t)} = [\opH_a,\avop{\rho(t)}] 
+ \avop{\opV(t)\rho(t)} - \avop{\rho(t)\opV(t)}.
\label{eq:tddm-av-1}
\end{equation}
Simple inspection of the above, given that generally  $\avop{\opV(t)\rho(t)} \neq \avop{\opV(t)} \avop{\rho(t)}$, reveals immediately that the key problem for the mean-value dynamics is the calculation of the (statistically) correlated products $\avop{\opV(t)\rho(t) }$ and $\avop{\rho(t) \opV(t)}$. These correlated products depend on the statistics of the random field, materialized in terms of the field's multitime-moments (coherences):
\begin{equation}
M_n(t_i,t_j, \cdots,t_n)  = \avop{\opV(t_i)\opV(t_{j})\cdots\opV(t_n)}.
\label{eq:moments}
\end{equation}
In the special case where the AC time and the system's evolution characteristic times are very different the solution of Eq.~(\ref{eq:tddm-av-1}) is drastically facilitated. In this work the point of departure is to assume the AC time, $\tau_c$, the shorter among the characteristic evolution times of $\rho(t)$; these may depend on the atomic system under question (energies, dipole matrix elements) and on other radiation field properties (e.g. frequency, maximum field strength). In relation to the latter, if $\opV(t)$ obeys Gaussian statistics the above moments factorize in products of the first two moments, the average mean, $\avop{\opV(t_i)}$ and the first-AC function, $\avop{\opV(t_i)\opV(t_j)}$; however, despite this reduction in terms of the lowest two moments, the higher-order moments may still strongly contribute in the system's EOMs; it is here, the multitime cumulant averages (or semi-invariants) enter as an alternative statistical machinery \cite{kubo:1962}. The cumulants exhibit what is called a 'cluster' behaviour in contrast to a 'factorization' property of moments for Gaussian processes [Eq.~(\ref{eq:gaussian_moments}]. Due to this, all cumulants, beyond the second, strictly vanish for Gaussian processes whereas for non-Gaussian ones they succesively decreases. So, the cumulants, have the appropriate behaviour to serve a twofold role: first, as successive terms of a perturbative expansion and second as a robust measure of the non-Gaussianity of the random field; any cumulant average of higher-order beyond the second it's due exclusively to the non-Gaussian statistical properties of the process. 

The cumulants of a random process is a combinatorial expansion of moments \cite{kubo:1962}. In the particular case (and of relevance here) of a zero mean value, $\avop{\opV(t)}=0$, the cumulants simplify considerably. For simplicity, assuming a scalar random processs and using a compact notation for the moments and the cumulants, $M_{ijk..n}=M_n(t_i,t_j,\cdots,t_n)$ and $C_{ijk..n}= C_n(t_i,t_j,\cdots,t_n)$, respectively, we have: 
\begin{align}
C_1 &= 0, 
\quad 
C_{12} = M_{12}, 
\quad 
C_{123} = M_{123}, 
\label{eq:cumulants14}
\\
\quad
C_{1234} &= M_{1234} - M_{12}M_{34} - M_{13}M_{24} - M_{14}M_{23}.
\nonumber
\end{align}
If, in addition, the field has Gaussian probability distribution function all the odd moments ($C_{123}= 0$) vanish while the even ones factorize in terms of the second moments ($M_{ij}$). For the fourth moment,
\[
 M_{1234} = M_{12}M_{34} + M_{13}M_{24} + M_{14}M_{23}, 
\]
with the immediate result $C_{1234} = 0$. Therefore for Gaussian processes we are left with the cumulants $C_1=0$ and $C_{12}=M_{12}$ as claimed. The higher cumulants vanish whereas the 
the higher moments are expressed in terms of the second moment $M_{ij}$, but do not vanish. 

In the next section we'll see that the density matrix equations are reduced to a nonlinear set of differential equations with respect to the field, containing terms proportional to the field and its square amplitude, $\calE(t)$ and $\calI(t)$. Thus we aim to a formulation of the EOMs in terms of cumulant averages which, as we argue, constitute a more appropriate perturbative expansion parameter than the moments (field's coherence functions).     
  
Concluding with this general discussion the reader who is particularly interested in various standpoints of the theory of random processes could advise related references \cite{wax:1954,vankampen:1976,fox:1978,gardiner:1985,vankampen:1992,lax:2006,loginov:1978}.

\subsection{Resonant ionization Density matrix equations}

Assuming a Fano picture ionization scheme as in Fig.~(\ref{fig:fig1}) \cite{lambropoulos:1981,2002:10}, we define by $\oprho(t)$ the atomic density matrix state we specify by $|g\rangle$, $|a\rangle$ the initial and the excited states with energies $E_g$ and $E_a$, respectively. In the density-matrix equation Eq.~(\ref{eq:tddm-av-1}) we take $\opV(t)=-\opd E(t)$ to represent the electric dipole interaction with  $\opd$ the atomic dipole operator and the external (linearly polarized) radiation field, modelled as 
\begin{equation}
E(t)= E_0(t) \cos(\omega t + \phi(t)).
\label{eq:field-real}
\end{equation}
Assume for a moment that no random fluctuations are present. For the near-resonant case the field-free evolution of the density matrix will vary as $E_a-E_g \sim \omega$, the same time scale as the periodic part of the field. In the presence of the field a lot slower time-variation on the density matrix elements of the system's state is superimposed $\sim |V_{ag}|\ll \omega$. We can take advantage of this difference and effectively remove the 'fast' oscillating contribution. We then have to deal with the field's envelope slow temporal variation ($\sim \tau_p \gg 2\pi/\omega$). Apart from the $\sim 1/|V_{ag}|$ time-scale the other characteristic time scales are determined by the autoionization decay rate $\sim 1/\Gamma_a$ and the maximum photoionization widths of the $|g\rangle$, $|a\rangle$ and $|c\rangle$ states. 

The field's fluctuations introduce into the dynamics another characteristic time scale, via its coherence time, $\tau_c$; this is something that it should properly be taken into account. The amplitude, $E_0(t)$ and the phase $\phi(t)$ are considered real random processes with known statistical properties. It is convenient to use the complex analytical field envelope, 
\begin{equation}
\calE(t) = \frac{1}{2}E_0(t)e^{\imath \phi(t)} = \calE_0(t) \epsilon(t), 
\label{eq:field}
\end{equation}
with $\calE_0(t)$ real and the fluctuations modeled by the $\epsilon(t)$ complex random processes, fully characterized via its statistical multitime moment functions. Accordingly, the intensity is defined as $\calI(t) = |\calE(t)|^2$. These slow-varying quantities may be considered as the time-average of the electric field and intensity over a time-interval containing several field cycles. 

With the field-free eigenvalue problem solved, the energies and the transition matrix element between the system's eigenstates can be calculated. Then we assume the density operator expanded over its Hamiltonian eigenstate basis. As usual, the diagonal terms $\rho_{ii}(t),~i=g,a,c,c^\prime$ provide the occupation probabilities for the system's eigenstates (\textit{populations}) while the off-diagonal ones $i\neq j, \rho_{ij}(t)$, (atomic \textit{coherences}) keep track of the phase relations among the various eigenstates.~Next we keep only the states essential to the dynamics, by effective elimination of the off-resonant continuum states (those states which are off-resonant to $E_g+\omega$ and $E_a+\omega$. Briefly, we transform to a slowly-varying representation with the diagonal elements left intact and the  coherences changed by a phase factor, $\sigma_{ag} = \rho_{ag}e^{\imath(E_a-E_g)t}$~\cite{Blum}. Following a standard procedure we arrive at a set of the density-matrix EOM parametrized in terms of the slowly-varying Rabi transition matrix element, ac-Stark shifts and ionization widths to the respective continua \cite{iop_book_lampros,georges:1980}: 
\begin{subequations}
\label{eq:tddms}
 \begin{flalign}
\dot{\sigma}_{gg} &= - \gamma_g \calI(t) \sigma_{gg} + 2\imag {D \calE (t) \sigma_{ag} },
\label{eq:gg}
\\
\dot{\sigma}_{aa} &= -\left[ \Gamma_a + \gamma_a \calI(t)\right] \sigma_{aa} -2\imag { D^*  \calE(t)\sigma_{ag} } 
\label{eq:aa}
\\
\dot{\sigma}_{ag} &=  
-\left[ \delta + \Delta \calI(t)\right]\sigma_{ag} 
+\imath D^* \calE^*(t)\sigma_{aa}
-\imath
D \calE^*(t)\sigma_{gg} 
\label{eq:ag}
\\
\dot{\sigma}_{cc} &
=
-\gamma_c \calI(t)  \sigma_{cc} 
- \dot{\sigma}_{gg}-\dot{\sigma}_{aa}
\\
\dot{\sigma}_{c^\prime c^\prime}
 &
 =  \gamma_a \calI(t) \sigma_{aa}
\label{eq:cpcp}
%
%
%
%
%
%
%
%
\end{flalign}
\end{subequations}
with $Im(z) \equiv (z-z^*)/2\imath$. The ionization widths, 
$\gamma_g,\gamma_a,\gamma_c$, are defined  
such that, $\gamma_g(t) = \gamma_g \calI(t)$ represents a slowly-varying photoionization width from $|g\rangle$ in the $|c\rangle$ continuum, $\gamma_a(t) = \gamma_a \calI(t)$ the corresponding (also slowly-varying) width from the resonant state $|a\rangle$ in the $|c^\prime\rangle$ continuum states and $\Gamma_a$ the autoionization width from $|a\rangle$ in the $|c\rangle$ continuum. For near-resonant laser frequencies this interatomic ionization channel interferes with the direct photoionization channel from the $|g\rangle$ state ($\gamma_g$); the latter interference, strongly dependent on the laser and atomic particulars, is reflected in the experimental observables e.g. ionization yield, photoelectron spectrum. Mathematically, this interference is represented in the dynamics equations by the complex Rabi transition matrix element, $\hat{D}$ and the complex peak detunings, $\hat{\Delta},\hat{\delta}$, the definition of which are given in the appendix A. 

Before proceeding to the main part of the present development we note that the first three equations which determine the bound-states dynamics are not directly coupled with the last two. The coupling with the continuum states is included indirectly in the ac-Stark shifts and the ionization widths but not directly with the continuum density matrix elements $\sigma_{cc}(t),\sigma_{c^\prime c^\prime}(t)$; so, effectively the continuum population dynamics does not affect the bound state's dynamics. On the other hand, asymmetrically, mere inspection of the remaining two equations for $\sigma_{cc}(t),\sigma_{c^\prime c^\prime}(t)$ shows that knowledge of the bound-state dynamics is essential. In our derivation below, we aim to follow a similar pattern, dictated from this remark;  We'll be first dealing with the bound state dynamics exclusively, and then we'll be applying the developed method to the continuum part, as well.


 
\tikzset{
pattern size/.store in=\mcSize, 
pattern size = 5pt,
pattern thickness/.store in=\mcThickness, 
pattern thickness = 0.3pt,
pattern radius/.store in=\mcRadius, 
pattern radius = 1pt}
\makeatletter
\pgfutil@ifundefined{pgf@pattern@name@_41vxeauyo lines}{
\pgfdeclarepatternformonly[\mcThickness,\mcSize]{_41vxeauyo}
{\pgfqpoint{0pt}{0pt}}
{\pgfpoint{\mcSize+\mcThickness}{\mcSize+\mcThickness}}
{\pgfpoint{\mcSize}{\mcSize}}
{\pgfsetcolor{\tikz@pattern@color}
\pgfsetlinewidth{\mcThickness}
\pgfpathmoveto{\pgfpointorigin}
\pgfpathlineto{\pgfpoint{\mcSize}{0}}
\pgfusepath{stroke}}}
\makeatother

 
\tikzset{
pattern size/.store in=\mcSize, 
pattern size = 5pt,
pattern thickness/.store in=\mcThickness, 
pattern thickness = 0.3pt,
pattern radius/.store in=\mcRadius, 
pattern radius = 1pt}
\makeatletter
\pgfutil@ifundefined{pgf@pattern@name@_lbmcwzdlq lines}{
\pgfdeclarepatternformonly[\mcThickness,\mcSize]{_lbmcwzdlq}
{\pgfqpoint{0pt}{0pt}}
{\pgfpoint{\mcSize+\mcThickness}{\mcSize+\mcThickness}}
{\pgfpoint{\mcSize}{\mcSize}}
{\pgfsetcolor{\tikz@pattern@color}
\pgfsetlinewidth{\mcThickness}
\pgfpathmoveto{\pgfpointorigin}
\pgfpathlineto{\pgfpoint{\mcSize}{0}}
\pgfusepath{stroke}}}
\makeatother

 
\tikzset{
pattern size/.store in=\mcSize, 
pattern size = 5pt,
pattern thickness/.store in=\mcThickness, 
pattern thickness = 0.3pt,
pattern radius/.store in=\mcRadius, 
pattern radius = 1pt}
\makeatletter
\pgfutil@ifundefined{pgf@pattern@name@_r4vh6mhd9 lines}{
\pgfdeclarepatternformonly[\mcThickness,\mcSize]{_r4vh6mhd9}
{\pgfqpoint{0pt}{0pt}}
{\pgfpoint{\mcSize+\mcThickness}{\mcSize+\mcThickness}}
{\pgfpoint{\mcSize}{\mcSize}}
{\pgfsetcolor{\tikz@pattern@color}
\pgfsetlinewidth{\mcThickness}
\pgfpathmoveto{\pgfpointorigin}
\pgfpathlineto{\pgfpoint{\mcSize}{0}}
\pgfusepath{stroke}}}
\makeatother
\tikzset{every picture/.style={line width=0.75pt}} 
\begin{figure}

\begin{tikzpicture}[x=0.75pt,y=0.75pt,yscale=-1,xscale=1]

\draw [line width=2.25]    (191.22,325.99) -- (298.83,326.81) ;
\draw [line width=2.25]    (188.11,224.21) -- (291.05,222.6) ;
\draw  [color={rgb, 255:red, 134; green, 40; blue, 40 }  ,draw opacity=1 ][pattern=_41vxeauyo,pattern size=6pt,pattern thickness=0.75pt,pattern radius=0pt, pattern color={rgb, 255:red, 122; green, 32; blue, 32}] (307.12,185.86) -- (401.51,185.86) -- (401.51,259.49) -- (307.12,259.49) -- cycle ;
\draw [color={rgb, 255:red, 126; green, 22; blue, 37 }  ,draw opacity=1 ][line width=2.25]    (249.05,325.35) -- (345.9,226.25) ;
\draw [shift={(349.39,222.67)}, rotate = 494.34] [fill={rgb, 255:red, 126; green, 22; blue, 37 }  ,fill opacity=1 ][line width=0.08]  [draw opacity=0] (14.29,-6.86) -- (0,0) -- (14.29,6.86) -- cycle    ;
\draw  [color={rgb, 255:red, 0; green, 0; blue, 0 }  ,draw opacity=1 ] (137.22,270.04) .. controls (136.67,272.26) and (136.15,274.36) .. (136.73,274.59) .. controls (137.32,274.82) and (138.85,273.11) .. (140.45,271.31) .. controls (142.05,269.5) and (143.58,267.79) .. (144.16,268.02) .. controls (144.75,268.25) and (144.23,270.35) .. (143.68,272.57) .. controls (143.13,274.78) and (142.61,276.89) .. (143.19,277.12) .. controls (143.78,277.34) and (145.31,275.63) .. (146.91,273.83) .. controls (148.51,272.03) and (150.04,270.31) .. (150.63,270.54) .. controls (151.21,270.77) and (150.69,272.88) .. (150.14,275.09) .. controls (149.59,277.3) and (149.07,279.41) .. (149.66,279.64) .. controls (150.24,279.87) and (151.77,278.15) .. (153.37,276.35) .. controls (154.97,274.55) and (156.5,272.84) .. (157.09,273.06) .. controls (157.67,273.29) and (157.15,275.4) .. (156.6,277.61) .. controls (156.05,279.83) and (155.53,281.93) .. (156.12,282.16) .. controls (156.7,282.39) and (158.23,280.68) .. (159.83,278.88) .. controls (161.44,277.07) and (162.96,275.36) .. (163.55,275.59) .. controls (164.13,275.82) and (163.61,277.92) .. (163.06,280.14) .. controls (162.51,282.35) and (161.99,284.46) .. (162.58,284.69) .. controls (163.16,284.91) and (164.69,283.2) .. (166.29,281.4) .. controls (167.9,279.6) and (169.43,277.88) .. (170.01,278.11) .. controls (170.59,278.34) and (170.07,280.45) .. (169.53,282.66) .. controls (168.98,284.87) and (168.46,286.98) .. (169.04,287.21) .. controls (169.62,287.44) and (171.15,285.72) .. (172.76,283.92) .. controls (174.36,282.12) and (175.89,280.41) .. (176.47,280.63) .. controls (177.06,280.86) and (176.54,282.97) .. (175.99,285.18) .. controls (175.44,287.4) and (174.92,289.5) .. (175.5,289.73) .. controls (176.09,289.96) and (177.61,288.25) .. (179.22,286.44) .. controls (180.82,284.64) and (182.35,282.93) .. (182.93,283.16) .. controls (183.52,283.39) and (183,285.49) .. (182.45,287.71) .. controls (181.9,289.92) and (181.38,292.03) .. (181.96,292.25) .. controls (182.55,292.48) and (184.08,290.77) .. (185.68,288.97) .. controls (187.28,287.17) and (188.81,285.45) .. (189.4,285.68) .. controls (189.98,285.91) and (189.46,288.02) .. (188.91,290.23) .. controls (188.36,292.44) and (187.84,294.55) .. (188.42,294.78) .. controls (188.61,294.85) and (188.89,294.73) .. (189.24,294.46) ;
\draw [color={rgb, 255:red, 0; green, 0; blue, 0 }  ,draw opacity=1 ][line width=1.5]    (128.95,265.83) -- (198.64,293.93) ;
\draw [shift={(201.42,295.05)}, rotate = 201.96] [color={rgb, 255:red, 0; green, 0; blue, 0 }  ,draw opacity=1 ][line width=1.5]    (14.21,-4.28) .. controls (9.04,-1.82) and (4.3,-0.39) .. (0,0) .. controls (4.3,0.39) and (9.04,1.82) .. (14.21,4.28)   ;
\draw [color={rgb, 255:red, 152; green, 11; blue, 28 }  ,draw opacity=1 ][line width=2.25]  [dash pattern={on 6.75pt off 4.5pt}]  (245.16,222.67) .. controls (257.29,193.96) and (310.84,168.35) .. (346.67,218.65) ;
\draw [shift={(349.39,222.67)}, rotate = 237.25] [fill={rgb, 255:red, 152; green, 11; blue, 28 }  ,fill opacity=1 ][line width=0.08]  [draw opacity=0] (14.29,-6.86) -- (0,0) -- (14.29,6.86) -- cycle    ;
\draw  [color={rgb, 255:red, 74; green, 144; blue, 226 }  ,draw opacity=1 ][pattern=_lbmcwzdlq,pattern size=6pt,pattern thickness=0.75pt,pattern radius=0pt, pattern color={rgb, 255:red, 25; green, 40; blue, 155}] (307.12,118.86) -- (401.34,118.86) -- (401.34,185.86) -- (307.12,185.86) -- cycle ;
\draw [color={rgb, 255:red, 74; green, 144; blue, 226 }  ,draw opacity=1 ][line width=2.25]    (380.18,258.78) -- (379.43,151.69) ;
\draw [shift={(379.4,146.69)}, rotate = 449.6] [fill={rgb, 255:red, 74; green, 144; blue, 226 }  ,fill opacity=1 ][line width=0.08]  [draw opacity=0] (14.29,-6.86) -- (0,0) -- (14.29,6.86) -- cycle    ;
\draw [color={rgb, 255:red, 0; green, 0; blue, 0 }  ,draw opacity=1 ][line width=1.5]    (245.16,222.67) .. controls (263.83,254.65) and (256.83,295.43) .. (249.05,325.35) ;
\draw [color={rgb, 255:red, 0; green, 0; blue, 0 }  ,draw opacity=1 ][line width=1.5]    (245.16,222.67) .. controls (228.82,251.01) and (229.6,298.34) .. (249.05,325.35) ;
\draw  [color={rgb, 255:red, 0; green, 0; blue, 0 }  ,draw opacity=1 ] (233.95,233.43) -- (243,229.02) -- (245.93,239.03) ;
\draw  [color={rgb, 255:red, 0; green, 0; blue, 0 }  ,draw opacity=1 ] (248.48,313.85) -- (247.29,323.7) -- (237.2,320.74) ;
\draw  [color={rgb, 255:red, 0; green, 0; blue, 0 }  ,draw opacity=1 ] (259.06,318.9) -- (250.77,322.78) -- (248.27,314.43) ;
\draw  [color={rgb, 255:red, 0; green, 0; blue, 0 }  ,draw opacity=1 ] (245.16,237.4) -- (248.19,227.59) -- (257.43,231.04) ;
\draw  [color={rgb, 255:red, 80; green, 89; blue, 71 }  ,draw opacity=1 ][pattern=_r4vh6mhd9,pattern size=6pt,pattern thickness=0.75pt,pattern radius=0pt, pattern color={rgb, 255:red, 132; green, 136; blue, 113}] (173.33,117.23) -- (268.49,117.23) -- (268.49,184.22) -- (173.33,184.22) -- cycle ;
\draw [line width=1.5]  [dash pattern={on 1.69pt off 2.76pt}]  (291.05,222.6) -- (364.34,221.02) ;
\draw [color={rgb, 255:red, 141; green, 150; blue, 131 }  ,draw opacity=1 ][line width=2.25]    (245.16,222.67) -- (244.43,150.77) ;
\draw [shift={(244.38,145.77)}, rotate = 449.42] [fill={rgb, 255:red, 141; green, 150; blue, 131 }  ,fill opacity=1 ][line width=0.08]  [draw opacity=0] (14.29,-6.86) -- (0,0) -- (14.29,6.86) -- cycle    ;

\draw (173.27,290.13) node  [rotate=-13.65] [align=left] {};
\draw (173.33,327.45) node  [font=\large,color={rgb, 255:red, 27; green, 37; blue, 14 }  ,opacity=1 ,xslant=-0.05] [align=left] {$\displaystyle |g\rangle $};
\draw (167.1,225.03) node  [font=\large,color={rgb, 255:red, 27; green, 37; blue, 14 }  ,opacity=1 ,xslant=-0.05] [align=left] {$\displaystyle |a\rangle $};
\draw (416.03,232.4) node  [font=\large,color={rgb, 255:red, 27; green, 37; blue, 14 }  ,opacity=1 ,rotate=-1.63,xslant=-0.05] [align=left] {$\displaystyle |c\rangle $};
\draw (247.23,274.29) node  [font=\large,color={rgb, 255:red, 65; green, 117; blue, 5 }  ,opacity=1 ,xslant=-0.05] [align=left] {\textcolor[rgb]{0,0,0}{D}\textcolor[rgb]{0,0,0}{$ $}};
\draw (311.01,272.84) node  [font=\large,color={rgb, 255:red, 27; green, 37; blue, 14 }  ,opacity=1 ,rotate=-315.14,xslant=-0.15] [align=left] {$\displaystyle \textcolor[rgb]{0.38,0.02,0.07}{\gamma }\textcolor[rgb]{0.38,0.02,0.07}{_{g}}\textcolor[rgb]{0.38,0.02,0.07}{\ }\textcolor[rgb]{0.38,0.02,0.07}{(}\textcolor[rgb]{0.38,0.02,0.07}{t}\textcolor[rgb]{0.38,0.02,0.07}{)}$};
\draw (290.12,175.22) node  [font=\large,color={rgb, 255:red, 27; green, 37; blue, 14 }  ,opacity=1 ,xslant=-0.05] [align=left] {$\displaystyle \textcolor[rgb]{0.55,0.05,0.11}{\Gamma }\textcolor[rgb]{0.55,0.05,0.11}{_{\alpha }}$};
\draw (169.33,268.25) node  [font=\large,color={rgb, 255:red, 0; green, 0; blue, 0 }  ,opacity=1 ,rotate=-28.09,xslant=-0.05] [align=left] {$\displaystyle \omega _{r}$};
\draw (159.13,292.16) node  [color={rgb, 255:red, 0; green, 0; blue, 0 }  ,opacity=1 ,rotate=-25.47] [align=left] {FEL};
\draw (431.36,147.86) node  [font=\large,color={rgb, 255:red, 27; green, 37; blue, 14 }  ,opacity=1 ,rotate=-1.72,xslant=-0.05] [align=left] {$\displaystyle |c''\rangle $};
\draw (353.13,167.95) node  [font=\large,color={rgb, 255:red, 74; green, 144; blue, 226 }  ,opacity=1 ,rotate=-358.79,xslant=-0.15] [align=left] {$\displaystyle \textcolor[rgb]{0.04,0.21,0.42}{\gamma }\textcolor[rgb]{0.04,0.21,0.42}{_{c}}\textcolor[rgb]{0.04,0.21,0.42}{\ }\textcolor[rgb]{0.04,0.21,0.42}{(}\textcolor[rgb]{0.04,0.21,0.42}{t}\textcolor[rgb]{0.04,0.21,0.42}{)}$};
\draw (158.55,151.4) node  [font=\large,color={rgb, 255:red, 27; green, 37; blue, 14 }  ,opacity=1 ,rotate=-1.72,xslant=-0.05] [align=left] {$\displaystyle |c'\rangle $};
\draw (218.78,195.49) node  [font=\large,color={rgb, 255:red, 74; green, 144; blue, 226 }  ,opacity=1 ,rotate=-358.79,xslant=-0.09] [align=left] {$\displaystyle \textcolor[rgb]{0.19,0.22,0.16}{\gamma _{a} \ ( t)}$};

\end{tikzpicture}

\caption{Sketch of the excitation/ionization near-resonant scheme with a FEL pulse.}
\label{fig:fig1}
\end{figure}
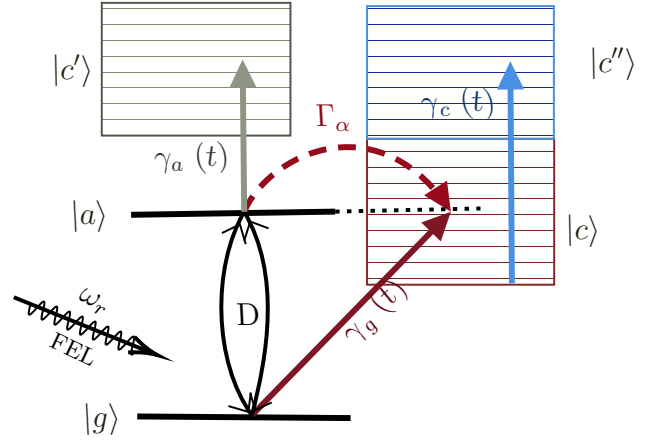

\subsection{Formal theory of averaging}
We start by considering only the set of Eqs.~(\ref{eq:gg})-(\ref{eq:ag}) and write the bound part of the Liouville equation: $\dot{\sigma}_b(t) = -\imath [\opH_b(t),\sigma_b(t)] \equiv \opL_b(t) \sigma(t)$ where $\sigma_b \equiv (\sigma_{gg}, \sigma_{aa},\sigma_{ag},\sigma_{ga})$,
with $\opL_b(t)$ temporally dependent on the fluctuating laser field, $\calE(t)$, and its intensity $\calI(t)$. At this point is sufficient to say that we assume the field to have zero ensemble mean average $\avop{\calE(t)}=0$. It is beneficial to split the random dynamics to a deterministic part, represented by the ensemble average of the Liouville operator, $\overline{L}_b(t) \equiv \avop{\opL_b(t)}$ and a random term,  $\widetilde{L}_b(t)$, with zero mean average, $\avop{\widetilde{L}_b(t)}=0$. This is done by splitting the intensity as $\calI(t) = \avop{\calI(t)} + \widetilde{\calI}(t)$ with, by construction, $\avop{\widetilde{\calI}(t)} = 0$. The bound part of the Liouville equation is now expressed by:
\begin{equation}
\dot{\sigma}_b(t) = 
[\overline{L}_b(t)  + \widetilde{L}_b(t)]\sigma_b(t),
\label{eq:rho_t}
\end{equation}
with initial conditions $\sigma_b(t_i) = (1,0,0,0)$. The mean-averaged part is a diagonal matrix,
\begin{equation}
\overline{L}_b(t)   = \opL_0 + \opL_2 \avop{\calI(t)},  
\label{eq:mean_L}
\end{equation}
while its random part is decomposed as:
\begin{equation}
\widetilde{L}_b(t) =  \opL_1\calE(t) +\opL_1^T\calE^*(t)+\opL_2\widetilde{\calI}(t). 
\label{eq:random_L}
\end{equation}
The explicit form of the ~$4\times4$ constant matrices, $\opL_i,~i=0-2$ and $\widetilde{L}_b(t)$ is relegated in the appendix A. The above decomposition results to a dynamics governed by non-commutative matrices,~$[\overline{L}_b(t), \widetilde{L}_b(t)] \neq 0)$,~($[\opL_1,\opL_i] \neq 0,~i=0,2 $ but $[\opL_0, \opL_2] = 0$ ); this is unfortunate, for if being otherwise the averaging procedure could be considerably simplified. Another point to note is that although the initial time is introduced as  $t_i$, at the end of the formal development it will be set to $t_i = -\infty$ (where the pulse is reasonably assumed negligible); this has the effect to render the averaged equations independent on the initial time, $t_i$; this is not just a practical matter but an essential step to turning properly a random differential equation to its averaged counterpart \cite{vankampen:1976}. The derived averaged equations will be valid as long as a transient period of the time scale of few $ \tau_c $.~The shortness of the coherence time results to fast decaying coherence functions for the field ensuring any dependence ('memory') on the initial time has been definitely lost;  a natural consequence of the randomness of the process in the course of time. 
  
Continuing with the formal steps of averaging Eq.~(\ref{eq:rho_t}) the deterministic part (mean average) of the dynamics is absorbed into a newly defined density matrix, through the transformation,
\begin{equation}
\sigma(t)  = e^{-\int_{t^\prime}^t d\tau^\prime \overline{L}_b(\tau^\prime)}\sigma_b(t) = \overline{U}_b(t,t_i) \sigma_b(t),
\label{eq:sigma_t} 
\end{equation}
with the (deterministic) evolution operator $\overline{U}_b(t,t_i)$ defined by inspection. Note here that $\avop{U_b(t,t_i)} \neq \overline{U}_b(t,t_i)$, albeit both sure functions of time.
Performing the transformation of Eq.~(\ref{eq:sigma_t}) in Eq. (\ref{eq:rho_t}) the effective (interaction-picture) time-evolution for the $\sigma(t)$ is,
\begin{equation}
\dot{\sigma}(t)   = \widetilde{L}(t) \sigma(t), \qquad  \sigma(t_i) = \sigma_b(t_i),
\label{eq:tddm_0}
\end{equation}
with the stochastic interaction matrix $\widetilde{L}(t)$ given by,
\begin{equation}
\widetilde{L}(t) = \overline{U}_b(t)\widetilde{L}_b(t)\overline{U}^\dagger_b(t).
\label{eq:l1_0}
\end{equation}
Given that $\overline{U}_b(t,t_i)$ is a deterministic evolution operator the mean value of the random interaction matrix, $\opL(t)$, vanish as well: 
\begin{equation}
\avop{\widetilde{L}(t)} = 
\overline{U}_b(t)\, \avop{\widetilde{L}_b(t)}\, \overline{U}^\dagger_b(t) = 0,
\label{eq:lzero}
\end{equation} 
since by construction $\avop{\widetilde{L}_b(t )} = 0$, as indicated earlier in the text. Temporarily, for convenience, the dependence on the initial time $t_i$ is suppressed by setting $\overline{U}_b(t) \equiv \overline{U}_b(t,t_i)$. 
Eq.~(\ref{eq:tddm_0}) represents a typical multiplicative system of stochastic differential equations, heavily studied over the years in various contexts \cite{vankampen:1976}. 


We chose to follow a perturbative approach by utilizing an infinite expansion solution of Eq.~(\ref{eq:tddm_0}). The formal solution is given by, 
\begin{align}
\sigma(t) & = \widetilde{U}(t)\sigma(t_i),
\label{eq:sigma-t-prime}
\end{align}
with the evolution operator, $\widetilde{U}(t)$, expressed in a chronologically ordered  format,~
$t > t_{1} > t_2 > \cdots > t_n$:
\begin{widetext}
\begin{equation}
\widetilde{U}(t) 
= \opOne + \sum_{n=1}^{\infty} \widetilde{U}_n(t) 
= \opOne + \sum_{n=1}^{\infty} 
\int_{t_i}^t dt_1 \int_{t_i}^{t_1} dt_2 \cdots \int^{t_{n-1}}_{t_i} dt_n \,\widetilde{L}(t_1)  \widetilde{L}(t_{2}) \cdots \widetilde{L}(t_n),  
\label{eq:dyson_v}
\end{equation}
\end{widetext}
where, the $\widetilde{U}_n(t)$ terms are defined by inspection. Since generally $[\widetilde{L}(t),\widetilde{L}(t^\prime)] \neq 0$  (for $t \neq t^\prime$) the chosen evaluation order of the operators in the integral matters. Formally, with $\sigma(t_i)$ a constant matrix, the average of Eq.~(\ref{eq:sigma-t-prime}) is calculated to, 
\begin{equation}
\avop{\sigma(t)} = \avop{\widetilde{U}(t)\sigma(t_i)} = \avop{\widetilde{U}(t)}\sigma(t_i)
\label{eq:evolution-moments}
\end{equation}
with the averaged evolution operator, $\avop{\widetilde{U}(t)}=\opOne + \sum_{n=1}U_n(t)$ as, 
\begin{flalign}
U_n(t)&=  \!\!\int_{t_i}^t \!\!dt_1 \int_{t_i}^{t_1} \!\!dt_2 \cdot \cdot \int^{t_{n-1}}_{t_i} \!\!dt_n\, M_n(t_1,t_2,\cdots,t_n),  
\label{eq:moments-1}
\end{flalign} 
and $M_n(t_1,t_2 \cdots t_n) $ as in Eq.~(\ref{eq:moments}) where $\opV(t)$ is replaced by $\tilde{L}(t)$, thus establishing the dependence on all field multitime moments (field's coherences). In this form the $\avop{\widetilde{U}(t)}$ operator as perturbative expansion may become impractical. In order to see this let's assume the various terms in the expansion at a particular time, say 't' and set the peak value of $\widetilde{L}(t)$ as $L_0 \sim D E_0, \gamma_g I_0, \gamma_a I_0, \Gamma_a$. The first term grows with time as $\sim tL_0$, the second term as $\sim t^2L_0^2$ and more generally a typical multi-time term grows as  $\sim (tL_0)^n$. For a well-behaved perturbative expansion we should require that $tL_0 \ll 1$. Therefore, regardless the peak strength $(\sim L_0)$ the value of an arbitrary term will depend crucially on the field's total duration, say $\tau_p$ through the combined parameter  $\tau_p L_0 \sim \tau_p L_0 \ll 1$. Taking into account a modest value $L_0 \sim 1$~a.u. the evolution operator of Eq.~(\ref{eq:evolution-moments}) is invalidated as a legitimate perturbative expansion at times much shorter $t >> 1$ a.u. ($> 0.024$ fs) than typical pulse duations of available FEL sources.

Based on the above observation, we aim to a perturbative expansion valid to arbitrarily long times via an expansion on cumulant averages instead on moment averages. Thus, we consider the series expansion the natural logarithm of $\avop{\widetilde{U}(t)}$ in terms of the cumulant averages for the $\widetilde{L}(t)$ operator:
\begin{align}
K(t) &= \ln \avop{\widetilde{U}(t)} = \opOne + \sum_{n=1} \kappa_n(t)
\label{eq:evolution-cumulants}
\\
\kappa_n(t) &= \!\int_{t_i}^t \!\!dt_1 \int_{t_i}^{t_1} \!\!dt_2 \cdot \cdot \int^{t_{n-1}}_{t_i} \!\!dt_n 
\; C_n(t_1,t_2,\cdots ,t_n). 
\nonumber
\end{align}
%
In the above expression a special prescription of the moments arises which leads to the so-called partially-ordered cumulants \cite{kubo:1962,vankampen:1992}. For our special choice, which leads to $\avop{\widetilde{L}(t)} = 0$, the lowest four cumulants are given by Eq.~(\ref{eq:cumulants14}). For the higher-order cumulants expressions, in terms of the moments $M_n(t_1,t_2,\cdots,t)$, may be found in the literature \cite{kubo:1962,vankampen:1992}. Again, it is important to note here that due to non-commutative property of  $\widetilde{L}(t)$ it is essential to keep track of the correct time-ordering for the partially-ordered cumulants.


With the above definitions the average of Eq.~(\ref{eq:sigma-t-prime}), using Eq.~(\ref{eq:evolution-cumulants}), can be written as,
\begin{equation}
\avop{\sigma(t)} = \avop{\widetilde{U}(t)} \sigma(t_i) = e^{K(t)} \sigma(t_i),
\label{eq:sigma_t_1}
\end{equation}
and by taking the time-derivative of its right-hand-side we obtain,
\begin{equation}
\frac{d}{dt} \left( e^{K(t) }\sigma(t_i) \right) = 
\hat{T}\left[\dot{K}(t)e^{K(t)}\right] \sigma(t_i)
=
\dot{K}(t)  \avop{\sigma(t)},
  \label{eq:dK_dt}
\end{equation}
where we used Eq.~(\ref{eq:sigma_t_1}) to replace $\sigma(t_i)$ and thus to revert back to $\avop{\sigma(t)}$.  In the above equation, the chronological operator $\hat{T}$ ensures the proper 
time-ordering ($t > t_{1} > t_2 > \cdots > t_n > t_i$). $\sigma(t_i)$ was moved outside the $\hat{T}$ bracket since it is placed at the rightmost part of the equation in accordance with the assumed time-ordering. 
$\dot{K}(t)$ is calculated by use of Eq.~(\ref{eq:evolution-cumulants}), to eventually arrive at the desired EOM for the mean ensemble average of the density matrix elements:
\begin{equation}
\frac{d}{dt} \avop{\sigma(t)} = \sum_{n=1}^{\infty}\dot{\kappa}_n(t) \avop{\sigma(t)},
\label{eq:tddm-averaged} 
\end{equation} 
with the time-derivatives of $\kappa_n(t)$ expressed in terms of the multitime cumulants by: 
\begin{align}
\dot{\kappa}_1(t) &= C_1(t), 
\label{eq:cumulants-n}
\\
\dot{\kappa}_n(t) &= 
\int_{t_i}^t \!\!dt_1 \int_{t_i}^{t_1} \!\!dt_2 \cdot \cdot \int^{t_{n-2}}_{t_i} \!\!dt_{n-1} 
\; C_n(t,t_1,\cdot \cdot,t_{n-1})
\nonumber 
\end{align}
with $t_0 = t$. The explicit calculation of the cumulants coefficients, $\dot{\kappa}_n(t)$ up to $n=4$ is given in the appendix in Eqs.~(\ref{eq:k1dot}),(\ref{eq:k2dot}),(\ref{eq:k3dot})(\ref{eq:k4dot}). 

The derivation of this formally exact expression did not necessitate any decorrelation approximation between the field's and $\sigma(t)$ fluctuations; neither any other particular statistical property such as stationarity or Gaussian-type multitime-correlations. Eq.~(\ref{eq:tddm-averaged}) holds for general random fields and may be used as a perturbative expansion as long as the AC time, $\tau_c$, is shorter than the other characteristic evolution times, a point which is discussed now. To show the argument, let's take the second term of the cumulant expansion, $\dot{\kappa}_2(t)$:
\begin{equation}
\dot{\kappa}_2(t) = \int_{t_i}^t dt_1  \;
\avop{\widetilde{L}(t)\widetilde{L}(t_1)}
\label{eq:k2-term}
\end{equation}
It shown later that this term is proportional to $\sim\avop{\calE(t)\calE^*(t_1)}$ and $\sim\avop{\calI(t)\calI(t_1)}- \avop{\calI(t)}\avop{\calI(t_1)}$. For values of $t$ and $t_1$ such that $|t-t_1| \gg \tau_c$ approximately $\avop{\calE(t)\calE^*(t_1)} \sim \avop{\calE(t)}\avop{\calE^*(t_1)} \sim 0 $ and $\avop{\calI(t)\calI(t_1)} \sim \avop{\calI(t)}\avop{\calI(t_1)}$. In this case the integrand of $\dot{\kappa}_2(t)$ is short-lived and approach negligible values so that effectively the integral is restricted in the time interval $t-t_1 \sim \tau_c$. Therefore it is concluded that $\dot{\kappa}_2(t) \sim \Gamma_a \tau_c, \gamma_g^2 \calI_0^2 \tau_c,\gamma^2_a \calI_0^2 \tau_c$ regardless the evaluation time $t$. This is in contrast with the evolution operator based on a moment expansion, Eq.~(\ref{eq:evolution-moments}), since the intensity term would still be contributing in the respective integral as  $\sim t \gamma_g^2 I_0^2, t \gamma_a^2 I_0^2$. Analogous conclusions are reached for the  higher cumulant terms which lead us to the necessary conditions for a converged perturbative expansion: 
\begin{equation}
\gamma_i I_0 \tau_c <  1, \quad i=g,a \qquad \text{and} \quad \Gamma_a \tau_c < 1,
\label{eq:conditions} 
\end{equation}
with $I_0$ the maximum peak intensity. Thus we have concluded that the convergence conditions of the averaged EOMs is that the field intensity should not reach ionization saturation conditions within the time scale of the AC time. 
%

A final point here is that our special choice of Eq.~(\ref{eq:rho_t}), which leads automatically to $\avop{\widetilde{L}(t)}= 0$, makes the two perturbative expansions for the averaged EOMs (in moments and in cumulant averages), equivalent up to the second order; this is no longer the case, though, as we proceed to higher orders.

\subsection{First two non-vanishing terms}

In the below we proceed by calculating the first two lowest-order of the cumulant expansion, $\dot{\kappa}_1(t)$ and $\kappa_2(t)$. For the $\dot{\kappa}_1(t)$ term, by use of Eqs.~(\ref{eq:cumulants14}), (\ref{eq:mean_L})-(\ref{eq:sigma_t}),(\ref{eq:l1_0}) and by recalling $\avop{\calE(t)}=\avop{\widetilde{\calI}(t)}=0$, we find $\dot{\kappa}_1(t)=0$. Using the same equations the $\dot{\kappa}_2(t)$ term of Eq.~(\ref{eq:k2-term}) is analyzed as, 
\begin{widetext}
\begin{equation}
\dot{\kappa}_2(t) =
\overline{U}_b(t) 
\int_{0}^{\infty} d\tau_1  
\avop{
\widetilde{L}_b(t)
e^{\int_{t-\tau_1}^{t} d\tau_1 \overline{L}_b(t,\tau_1)} 
\widetilde{L}_b(t-\tau_1)} 
e^{-\int_{t-\tau_1}^{t} d\tau_1 \overline{L}_b(t,\tau_1)}
 \overline{U}^{\dagger}_b(t). 
\label{eq:k1}
\end{equation}
\end{widetext}
To arrive at the last line in Eqs.~(\ref{eq:sigma_t}) and (\ref{eq:l1_0}) the initial time was set $t_i=-\infty$, followed by a change to the relative time variable $\tau_1 = t-t_1$. The last change has effectively removed the dependence of the equation on the initial time, as desired; as discussed a meaningful average law should lose any memory at times much longer than the fluctuation's correlation time. Keeping on with the evaluation, we take advantage of the shortness of the correlation-time relative to the average intensity's slow variation and express the integral over the $\overline{L}_b(t,\tau_1)$ more conveniently:
\begin{equation}
\int_{t^\prime}^t d\tau^\prime \overline{L}_b(\tau^\prime)
=
\opL_0(t-t^\prime) + \opL_2 \avop{W(t,t^\prime)},
\end{equation}
where $\avop{W(t,t^\prime)}$ is the mean energy contained in the time interval $\tau$:
\begin{equation}
\avop{W(t,t^\prime)} =  \int_{t^\prime}^t d\tau^\prime \avop{\calI(\tau^\prime)} 
\label{eq:energy-dt}
\end{equation}

This term depends on the static part of the Hamiltonian and on the field's mean energy. A change of variables to $\tau_1=t-t^\prime > 0$, gives,
\begin{align*}
\int_{t-\tau_1}^{t} d\tau_1 \overline{L}_b(t,\tau_1) 
&= \opL_0 \tau_1 +  \opL_2 \int_{t-\tau_1}^t dt^{\dprime} \avop{\calI(t^{\dprime} )} 
\\
& \simeq (\opL_0 + L_2\avop{\calI(t)}) \tau_1 
=
\overline{L}_b(t) \tau_1.
\end{align*}
The approximate replacement in the second line is related with the slow variation of the field-envelope. More specifically, since the field's AC functions decay with a rate $\sim 1/\tau_c$ practically the integral in Eq.~(\ref{eq:k1}), takes its main contributions for times $\tau_1$ in the interval $[t-\tau_c,t]$; therefore the slowly varying $\calI(t)$ is practically constant in this time interval. In quantitative terms a Taylor expansion of $\avop{I(t^\dprime=t-\tau)}$ gives, 
\begin{align*}
\int_{t-\tau_1}^t dt^{\dprime} \avop{\calI(t^{\dprime})} &= 
\int^{\tau_1}_{0} d\tau
(
\avop{\calI(t)}\tau - \tau \avop{\dot{\calI}(t)}  + \cdots 
)  
\\
&\simeq  \avop{\calI(t)} \tau_1 - \mathcal{O} \Big(\tau_c\avop{\dot{\calI}(t)}\Big)\tau_1/2, 
\end{align*}
with the leading error term replaced by $\tau_1 \avop{\dot{\calI}(t)} \simeq   \tau_c \avop{\dot{\calI}(t)} \ll 1 $.

With the above in mind, now we transform back to the original EOMs $\sigma(t)$ using Eq.~(\ref{eq:sigma_t}) to arrive at,
\begin{equation}
\avop{\dot{\sigma}_b(t)} = \left[\overline{L}_b(t)  
+ \!\int_0^\infty \!\!d\tau_1 \calK_2(t,t-\tau_1)
\right] \avop{\sigma_b(t)} 
\label{eq:sigma-k2}
\end{equation}
with the extra term $\calK_2(t,t_i)$ clearly generated from the averaging process. 
The term inside the integrand is non-vanishing for times of the order of the field's correlation time, $\tau_c$, since it depends on the first two AC functions. More specifically, direct substitution of Eqs.~(\ref{eq:mean_L}) and (\ref{eq:random_L}) gives,
\begin{align}
\calK_2(t,t-\tau_1)
&=
\opL_1
e^{\overline{L}_b(t)\tau_1} 
\opL_1^{T}
e^{ -\overline{L}_b(t)\tau_1}
\avop{\calE(t)\calE^*(t-\tau_1)} 
\nonumber
\\
&+ 
\opL_1^T
e^{ \overline{L}_b(t)\tau_1} 
\opL_1
e^{-\overline{L}_b(t)\tau_1}
\avop{\calE^*(t)\calE(t-\tau_1)}
\nonumber\\
&+ \opL_2^2 
\avop{\calI(t)\calI(t-\tau_1)} 
\label{eq:k1-c1}
\end{align}
Explicit substitution in Eqs.~(\ref{eq:k1-c1}) and (\ref{eq:sigma-k2}) of $\opL_i~i=0,1,2$ 
[Eqs.~(\ref{eq:L0}), (\ref{eq:L1}) and (\ref{eq:L2})] leads to our main expressions for the averaged EOMs of the bound states:
\begin{widetext}
\begin{subequations}
\label{eq:avg-b}
\begin{align}
\frac{d}{dt}\avop{\sigma_{gg}(t)}& = 
-
\left[
 \gamma_g \avop{\calI(t)} -  \gamma^2_g S_{2t}(0)   
 + 2 Re\{ D^2 S_{1t}(\hat{\delta}_+)\}
 \right] 
 \avop{\sigma_{gg}(t)} 
+
 2 |D|^2 Re\{S_{1t}(\hat{\delta}_-)\} 
\avop{\sigma_{aa}(t)}
 \\
\frac{d}{dt}\avop{\sigma_{aa}(t)}& = 
-
\left[\Gamma_a +
\gamma_a 
\avop{\calI(t)} -  \gamma^2_a  S_{2t}(0) 
+
2Re\{D^{*^2} S_{1t}(\hat{\delta}_-)\}
\right]
\avop{\sigma_{aa}(t)}
+
 2  |D|^2 Re\{S_{1t}(\hat{\delta}_+)\}  
\avop{\sigma_{gg}(t)}
\\
\frac{d}{dt}\avop{\sigma_{ag}(t)}& =
-\left[ 
\delta + \Delta \avop{\calI(t)} - \Delta^2 
S_{2t}(0)
+
\left\{
D^2\calS_{1t}(-\hat{\delta}_+) + D^{*^2}\calS_{1t}(-\hat{\delta}_-) 
\right\}
\right]
\avop{\sigma_{ag}(t)},
\end{align}
\end{subequations}
\end{widetext}
where the averaged dynamic detunings, $\hat{\delta}_{\pm} \equiv \hat{\delta}_\pm(t)$, are defined by,
\begin{equation}
\hat{\delta}_{\pm}(t) = \imath \left[\delta_0 + (s_a-s_g)\avop{\calI(t)} \right]
\pm 
\frac{\Gamma_a + (\gamma_a-\gamma_g) \avop{\calI(t)}}{2}
\label{eq:dpm_t}
\end{equation}
%
%
The coefficients in Eqs.~(\ref{eq:avg-b}) are expressed as the Laplace transforms of the AC functions of the field and the intensity, respectively:
\begin{align}
\calS_{1t}(\hat{\delta}) &=  \int_{0}^{\infty} d\tau\; 
\avop{\calE(t)\calE^{^*}(t-\tau)} \; e^{-\hat{\delta}\tau}, 
\label{eq:s1}
\\
\calS_{2t}(0) &=  \int_{0}^{\infty} d\tau\; 
 \left( \avop{\calI(t)\calI(t-\tau)}   - \avop{\calI(t)}\; \avop{\calI(t-\tau)}\right) .
 \label{eq:s2}
\end{align}
The complex quantity $S_{1t}(\hat{\delta})$ is closely related with a time-dependent frequency spectrum and for such long pulses, where stationarity conditions of the field's statistically properties are reached, approach the field's power spectrum via the Wiener–Khinchin theorem \cite{Krinsky1}. Under the later (stationarity) assumption it can also be shown that the $S_{2t}(0)$ is proportional to the field's average energy standard deviation, $\avop{\Delta W^2(t)}$. 
 Eqs.~(\ref{eq:avg-b}) are the result of the lowest-order non-vanishing approximation of an infinite term expansion and have general applicability with respect to the field's fluctuations on the proviso of the conditions set by Eq.~(\ref{eq:conditions}); as such there is an associated upper limit in the intensity for the condition to be fulfilled.

Explicit, analytical, expressions for the $S_{1t},S_{2t}$ are given later in the text for the case where the field is considered non-stationary, Gaussian and square-exponentially time correlated coherence functions.

\subsection{Equations for the continuum populations}

Now we turn to the derivation of the averaged populations of the continuum part of the density matrix, namely, the time evolution law for $\avop{\sigma_{cc}(t)}$ and $\avop{\sigma_{c^\prime c^\prime}(t)}$. These equations constitute an inhomogeneous set of stochastic equations dependent on the statistically correlated terms, $\calI(t)$, $\sigma_{gg}(t)$ and $\sigma_{aa}(t)$:
\begin{subequations}
\begin{align}
\dot{\sigma}_{cc} &
=
-\gamma_c \calI(t)  \sigma_{cc}  - \dot{\sigma}_{gg} - \dot{\sigma}_{aa}
\label{eq:cc-2}
\\
\dot{\sigma}_{cc} & = \gamma_a \calI(t) \sigma_{aa}.
\label{eq:cpcp-2}
\end{align}
\end{subequations}
The statistics of $\calI(t)$ is given from the outset whereas this is not the case for the random populations, $\sigma_{gg}(t)$, $\sigma_{aa}(t)$; their statistical properties are determined by the atomic system's response. Various methods are available to derive the averaging form of the above equations but since here we focus on the first non-vanishing terms for arbitrary stochastic fields we may repeat the method applied in the bound part of the density matrix. For this, the first task is to bring the continuum equations in the form of Eq.~(\ref{eq:rho_t}). We achieve this by augmenting the set of the continuum equations to include the $\dot{\sigma}_{gg}, \dot{\sigma}_{aa}$ and define $\sigma_c(t) = ( \sigma_{cc}(t),\sigma_{c^\prime c^\prime}(t), 1)$ with $p_b(t) = \sigma_{gg}(t) + \sigma_{aa}(t)$. Now we can write,
\begin{equation}
\dot{\sigma}_c(t) 
=
\left[ \overline{L}_c(t) + \widetilde{L}_c(t) \right]
\sigma_c(t), 
\end{equation}
with $\sigma_c(0) = (0,0,1)$. The mean-average and the random parts, $\overline{L}(t)$, $\widetilde{L}(t)$, are given by ,
\begin{align} 
\overline{L}_c(t) &=
\begin{pmatrix}
-\gamma_{c}\avop{\calI(t)}  & 0    & 0\\ 
 0 & \gamma_a \avop{\calI(t)} &  0 \\ 
 0 & 0 & 0
\end{pmatrix},
\\
\widetilde{L}_c(t) &=
\begin{pmatrix}
-\gamma_{cc}\widetilde{\calI}(t)  & 0    & \dot{p}_b(t)\\ 
 0 & \gamma_a \widetilde{\calI}(t) &  0 \\ 
 0 & 0 & 0
\end{pmatrix}.
\end{align}
Repeating the steps described in the previous sections and keeping only the first two terms of the perturbative expansion we arrive to an equation analogous to Eq.~(\ref{eq:sigma-k2}). The end result for the averaged populations is,
\begin{widetext}
\begin{subequations}
\label{eq:avg-c}
\begin{align}
\avop{\dot{\sigma}_{cc}(t)}& = 
-  
\gamma_c \avop{\calI(t)}
\avop{\sigma_{cc}} 
- \avop{\dot{\sigma}_{gg}} - \avop{\dot{\sigma}_{aa}}
+\gamma_c S_{2t}(0)
\left[
\gamma_c \avop{\sigma_{cc}}
-
\gamma_g \avop{\sigma_{gg}}
 -
\gamma_a \avop{\sigma_{aa}}
\right]
\\
\avop{\dot{\sigma}_{c^\prime c^\prime}(t)}& = +
\left[
 \gamma_a \bar{\calI}(t) +  \gamma^2_a S_{2t}(0)  
\right] 
\avop{\sigma_{aa}(t)} 
\end{align}
\end{subequations}
\end{widetext}
%
%
%

\section{The case of a Gaussian random field}

Having established the approximate averaged equations we now proceed to further specialization for the important case of pulsed fields obeying Gaussian statistics. 
The field is assumed with an envelope of the form: 
\begin{equation}
\calE_0(t)  = \calE_0 e^{-\hat{\chi} t^2/2\tau_p^2},
\label{eq:gaussian-envelope}
\end{equation}
with $\hat{\chi}$ complex, to allow for chirped pulses \cite{saldin_book:2000,Krinsky1}. This 
choice specify the statistically averaged intensity via Eq.~(\ref{eq:field}) as, 
\begin{equation}
\avop{\calI(t)} = 
 \avop{|\calE(t)|^{{^2}}}= \calE_0^2 e^{-t^2/\tau^2_p}.
 \label{eq:mean-intensity}
\end{equation} 
Note that the FWHM duration of the mean intensity is related with $\tau_p$ by $\tau_{FWHM} = \tau_p \sqrt{4\ln 2}$. The reader should not be confused and associate fields with this particular choice of a Gaussian envelope with fields obeying Gaussian statistics; it is unrelated with the Gaussian statistical properties of the field and the square-exponential decay form of the AC functions. We could have chosen any other analytical form for the envelope of the mean intensity to model the actual pulse.  
 
In view of  Eq.~(\ref{eq:field}), the field's fluctuations are modelled through the stationary random field, 
$\epsilon(t)$, obeying complex Gaussian ensemble-statistics \cite{goodman:1985}, 
\begin{equation}
\avop{\epsilon(t)} = \avop{\epsilon(t)\epsilon(t^\prime)} =  
\avop{\epsilon^{*}(t)\epsilon^{*}(t^\prime)} = 0.
\end{equation}
This results to the non-stationary first-order field coherence,
\begin{equation}
\avop{\calE(t)\calE^{^*}(t^\prime)} = \calE_0(t)\calE^*_0(t^\prime)\avop{ \epsilon(t) \epsilon^{*}(t^\prime) }.
\label{eq:c1-field} 
\end{equation} 
Under these conditions the first-order coherence function of the intensity can be related to the first-order coherence of the field:
\begin{equation}
\avop{\calI(t)\calI(t')} = \avop{\calI(t)}\;\avop{\calI(t^\prime)}+ 
|\avop{\calE(t)\calE^{^*}(t^\prime)}|^2, 
\label{eq:c1-intensity-gaussian}
\end{equation}
This gives for the random part of the intensity:
\begin{equation}
\avop{\widetilde{\calI}(t)\widetilde{\calI}(t')} = |\avop{\calE(t)\calE^{^*}(t^\prime)}|^2,
\label{eq:c1-field} 
\end{equation}
which effectively results to expressing $S_{1t}(\pm\hat{\delta}_{\pm})$ and $S_{2t}(0)$ exclusively in terms of the first-order field's coherence function $\avop{\calE(t)\calE^{^*}(t^\prime)}$.  

\paragraph{Square-exponentially correlated 1st-order coherence.} 
This consists to assume the below dependence:  
\begin{equation}
 \avop{\epsilon(t)\epsilon^*(t^\prime)} = e^{-(t-t^\prime)^2/2\tau^2_c}. 
 \label{eq:c1-gauss}
\end{equation}
Regarded as a sufficiently good approximation to actual FEL fields, this model has been studied in detail from the viewpoint of a shot-noise random field resulting to a random intensity with exponential probability distribution law and fluctuating energy, $W(t,t_i)$, of Gamma-distribution \cite{Rice1944,Krinsky1,kim:2017}.
The required integrations are performed analytically to provide the closed-form expressions,
\begin{subequations}
\begin{align}
S_{1t}(\hat{\delta}) &= \tau_{coh}\sqrt{\pi} \avop{I(t)} 
\hat{\lambda} w(\hat{\lambda}(\frac{\hat{\chi}^*t}{M\tau_p} - \hat{\delta}\tau_{coh}) )
\label{eq:s110}
\\
S_{2t}(0) 
&= 
 \tau_{coh} \sqrt{\pi}
\frac{\avop{I(t)}^2}{2}\; w(\frac{t}{M\tau_p})
\label{eq:s220}
\\
M &= \frac{\tau_p}{\tau_{coh}}, 
\qquad
\tau_{coh} = \frac{\tau_p \tau_c}{\sqrt{\tau_p^2 + \tau_c^2}} 
\\
\hat{\chi} &= 1 - \imath k, \quad \hat{\lambda} = \frac{1}{\sqrt{2(1+\imath k/M^2)}}
\label{eq:tcoh}
\end{align}
\end{subequations}
where $w(z) = e^{z^2}(1+\text{erf}(z))$, $\text{erf}(z)$ to be the error function and 
$k = \frac{1}{\sqrt{3}}$ for FEL pulses. These expressions are very closely related with the Voight profile functions, originating from the convolution of a Gaussian and Lorentzian spectra. The Gaussian character of the resulting spectrum is exclusively due to the field's properties, namely the envelope's mean-value time-dependence ($\sim\tau_p$) and coherence time ($\sim\tau_c$). On the other hand the Lorentzian character is due to near-resonance effects and thus includes information about the atomic system and its response to the atomic field via the mean-average detunings $\hat{\delta}_{\pm}(t)$; as the latter quantity includes information only about the mean intensity, $\avop{I(t)}$ it may be infered that the Lorentzian character of the lineshape is not influenced by the field's fluctuations. The $M$ parameter is a measure of the number of random 'spikes' during the pulse and of the strength of the fluctuations as its inverse approximates $\avop{(\Delta W)^2}/\avop{W}^2$ where $\avop{(\Delta W)^2}$ is the standard deviation of $W(t)$. Finally the $\tau_{coh}$ is known as the field's \textit{coherence time} and may be alternatively defined via the normalized field's coherence function. More detail on this and the various field parameters can be found in \cite{Krinsky1}.

For longer pulses ($M \gg 1 ~(\tau_p \gg \tau_c)$), $S_{1t},S_{2t}$ take a simpler form; at times $t\sim M \tau_p$ the mean intensity $\avop{I(t)}$ is negligible (decays as $e^{-2(t/\tau_p)^2}$) and we may consider the earlier times $t\sim \tau_p$ where the approximation $w(t/M\tau_p + b) \sim w(0+b)$ can be adopted. For such pulses the time-dependence and the frequency dependence in $S_{1t}(\hat{\delta})$ 
factorize as below: 
\begin{subequations}
\begin{align}
S_{1t}(\hat{\delta}) &\simeq \tau_{coh}\sqrt{\pi}\frac{\avop{\calI(t)}}{\sqrt{2}} 
\;
w(\frac{-\hat{\delta}\tau_{coh}}{\sqrt{2}})
\; 
\label{eq:s1ta}
\\
S_{2}(0) & \simeq  \tau_{coh} \sqrt{\pi} \frac{\avop{\calI(t)}^2}{2} \; .
\label{eq:s20a}
\end{align}
\end{subequations}
From Eq.~(\ref{eq:s1ta}) it is easier the clarify the role of the various parameters in the line-shape profile; we see that in the argument of Eq.~(\ref{eq:s1ta}) the smooth time-dependence on the mean intensity is still present as seen from Eq.~(\ref{eq:dpm_t}). So $S_{1t}$ depends on the atomic parameters and the smooth mean intensity envelope and the field's coherence time, $\tau_{coh}$ (note that for $M\gg 1, \; \tau_{coh}\to \tau_c$).  It's not difficult to see that with decreasing $\tau_{coh}$ the lineshape takes a Lorentzian shape regardless the field's intensity. In such conditions, the shape of the Lorentzian profile will not contain any direct influence of the field's fluctuations statistics; this will depend on the parameters included in the $\hat{\delta}_\pm$ mean detunings.
  
 
\paragraph{Exponentially correlated 1st-order coherence.} 

Now we assume the AC coherence function to exhibit exponential correlation 
behaviour of the type:
\begin{equation}
 \avop{\epsilon(t)\epsilon^*(t^\prime)} = e^{-|t-t^\prime|/2\tau_c}
\label{eq:c1-exponential}
\end{equation}
For simplicity we'll consider the case of non-chirped pulses and take $\hat{\chi}=1$ in Eq.~(\ref{eq:gaussian-envelope}) (the case of chirped pulse can be easily incorporated, when needed). In this case we obtain for $S_{1t}$ and $S_{2t}$,
\begin{align}
    S_{1t}(\delta) =& \tau_p\sqrt{\pi} \frac{\avop{\calI(t)}}{\sqrt{2}} w(\frac{\tau_\delta}{\sqrt{2}}),
\\
S_{2t}(0) =& \tau_p \sqrt{\pi} \frac{\avop{\calI(t)}^2}{2} w(\tau(0)),
\\
   \tau_\delta &= \frac{t}{\tau_p} - (\frac{1}{2\tau_c} +\delta) \tau_p  
\end{align}
thus, still showing a Voight profile shape. Using the asymptotic value $w(z \gg 1) \to 1/z\sqrt{\pi}$, we find for the corresponding approximate expressions for longer pulses ($\tau_p \gg \tau_c$):
\begin{align}
    S_{1t}(\delta) =& \avop{\calI(t)}\frac{2\tau_c}{1+2\delta\tau_c}\\
    S_{2t}(0) =& \avop{\calI(t)}^2\tau_c,
\end{align}
show a clearly a Lorentzian-like shape.

\begin{table}
 \begin{tabular}{|c c|} 
 \hline
 Parameters &  Values (a.u.)\\ 
 \hline
 $(E_a, q_a)$  & (2.211, -2.733) \\ 
 \hline
  $d_{ga}$  &    0.0358 \\
 \hline
 $\Gamma_{\alpha}$ &    0.00137  \\
 \hline
  $\gamma_g$ &  0.494   \\
 \hline
  $\gamma_c$  &   0.426  \\
 \hline
 \end{tabular}
  \caption{Atomic parameters in a.u.~used for the He$(2s2p)$ AIS resonance; The values for
  the effective matrix element $d_{ga}$ is calculated by $4|d_{ga} |^2 = q_a^2\; \Gamma_{\alpha}\gamma_g$. 
  }
  \label{tab:param_table_helium}
\end{table}
\section{Results and Discussion}

In this section we apply the averaged equations for the resonance ionization of Helium via the 2s2p $^1$P AIS located circa $ 60.154$~eV above its ground state. 
First, we compare the yields obtained from the averaged equations with those obtained from a MC set of calculations in order to establish numerically the range of validity of the approximation. Then we examine the role of the field's- and the intensity's- coherence functions on the ionization process. Finally, we proceed to compare the effects the field's-correlation time dependence on the AIS lineshape. The specific set of the atomic parameters in the following calculations are given in Table~(\ref{tab:param_table_helium}). These values have been calculated some time earlier with 
a well established atomic calculation package \cite{2002:10,2003:cpc}  

\subsection{Comparison with the Monte Carlo calculations}
\begin{figure*}
\begin{tabular}{c c}
  \includegraphics[width=65mm,scale=0.5, angle=270]{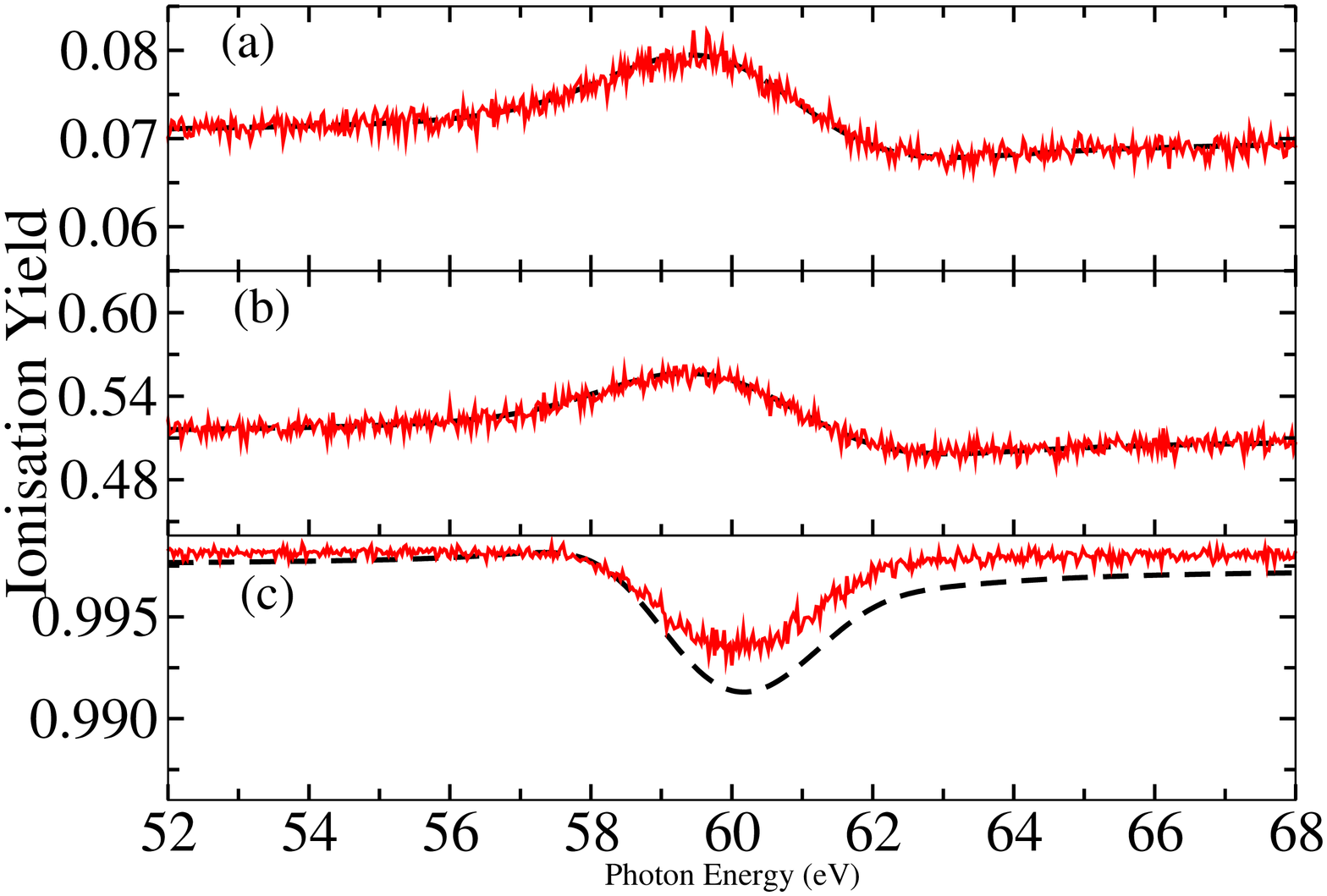} &
    \includegraphics[width=65mm,scale=0.5, angle=270]{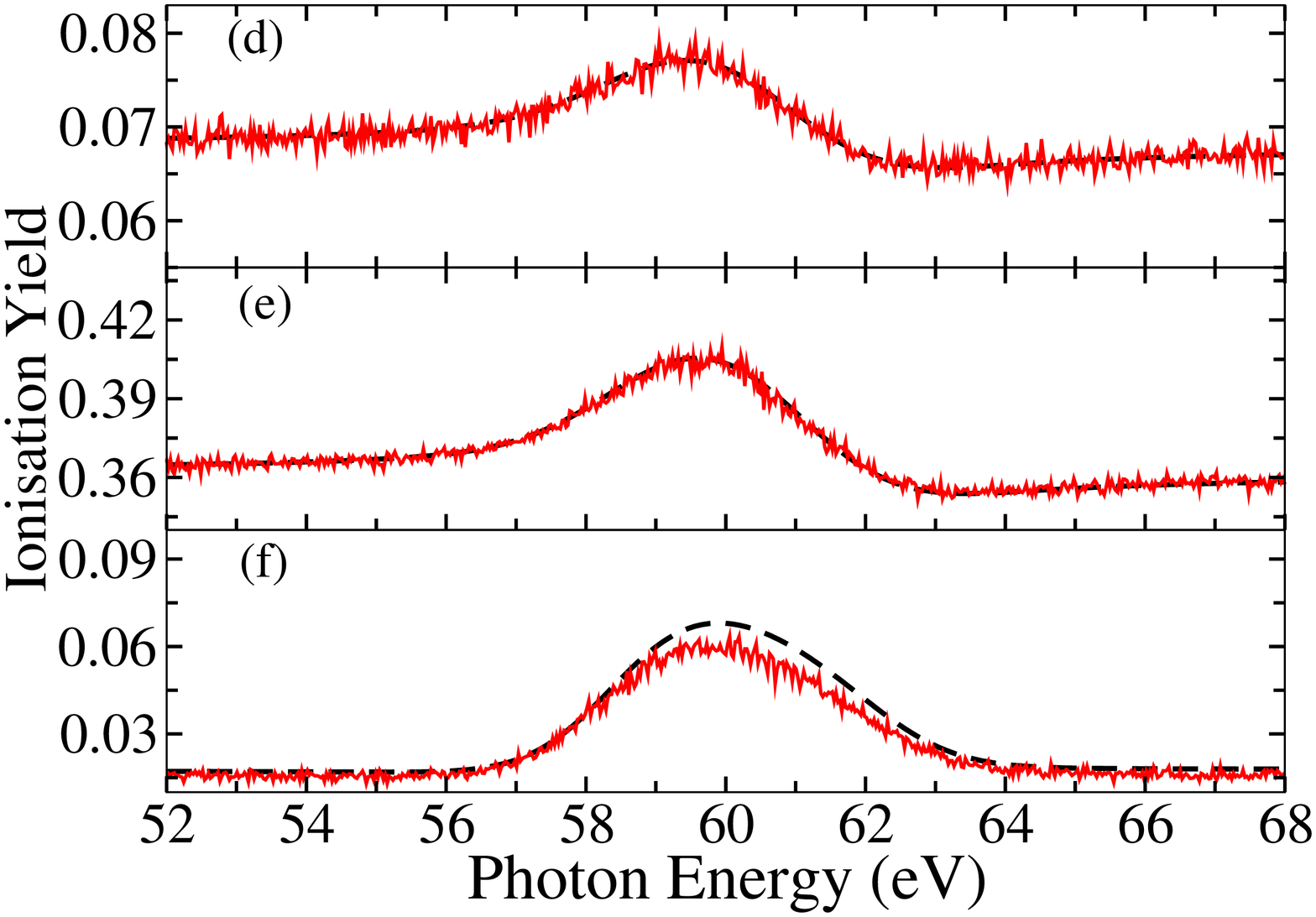} 
\end{tabular}
\caption{Comparison of ionisation yields obtained from th Monte Carlo method (red solid curve) and averaged density matrix equations (black dashed curve) for peak intensities of $10^{13}$ W/cm$^2$,  $10^{14}$ W/cm$^2$ and $10^{15}$ W/cm$^2$ (top to bottom respectively). $\gamma_c=0$ a.u. for  (a), (b), (c) and $\gamma_c=0.426$ a.u. for (d), (e), (f). Other parameters are $\tau_p = 7$ fs, $\tau_c = 0.5$ fs ($M = 14$).}
\label{fig:avg_mc_setA}
\end{figure*}

Using the averaging method, not only the computational demands are significantly reduced but a clearer insight of the ionization dynamics under stochastic fields may be obtained. To investigate the validity of the equations we compare the results obtained with the averaged equations against those obtained via a 
Monte Carlo (MC) approach. In the latter method the original set of the DM equations [Eq.~(\ref{eq:tddms})] is solved for a large number of (random) realizations of the field. At the end, the results are collected and are accordingly averaged. In principle, when the number of the MC trials goes to infinity (in practice this number depends on the problem) one should expect the results between the two calculational methods (MC and averaged equations) to coincide to a high degree. 

In Fig.~(\ref{fig:avg_mc_setA}), we plot the ionisation yields, obtained from the MC simulation and the averaged equations Eqns.~(\ref{eq:avg-b}),(\ref{eq:avg-c}), for different peak intensities of the pulse. The irregular behaviour of the MC yield is the result of the inherent randomness present in the FEL field, which diminishes with the total number of MC trials increasing; As the computational time increases, we chose 300 simulations to calculate the MC average, as the tendency is clearly evident. For the following calculations, we set $\tau_p = 7$~fs, $\tau_c = 0.5$ fs and $M = 14$.


The plots (a),(b) and (c) of this figure show the single ionization yield without including the subsequent ionization step from He$^+$ and without considering ionization directly from the AIS state. For this, we have assumed $\gamma_c = \gamma_a = 0$ and calculated for three different peak intensities, $10^{13}$ W/cm$^2$,~$10^{14}$ W/cm$^2$ and $10^{15}$ W/cm$^2$. Simple inspection of the plots shows that for the higher intensity the agreement between the two methods is less satisfactory, especially in the proximity of the resonance. For this highest peak intensity the perturbative terms due to the intensity correlation take the peak values $\gamma_g I_0 \tau_c \sim 0.3$ ($\Gamma_a \tau_c \sim 0.028$) in fair consistency with the conditions of Eq.~(\ref{eq:conditions}). The cumulant perturbative expansion was truncated in the second order which includes the first two coherences functions of the field. So, at this peak intensity the next perturbative term should be included in the averaged equations, but for the intensities currently available at this wavelength and the current purposes of this work it's not vital to include the next term of the expansion; however we'll be returning on this matter later in the conclusion section.

\begin{figure}

 \includegraphics[width=65mm,scale=0.5]{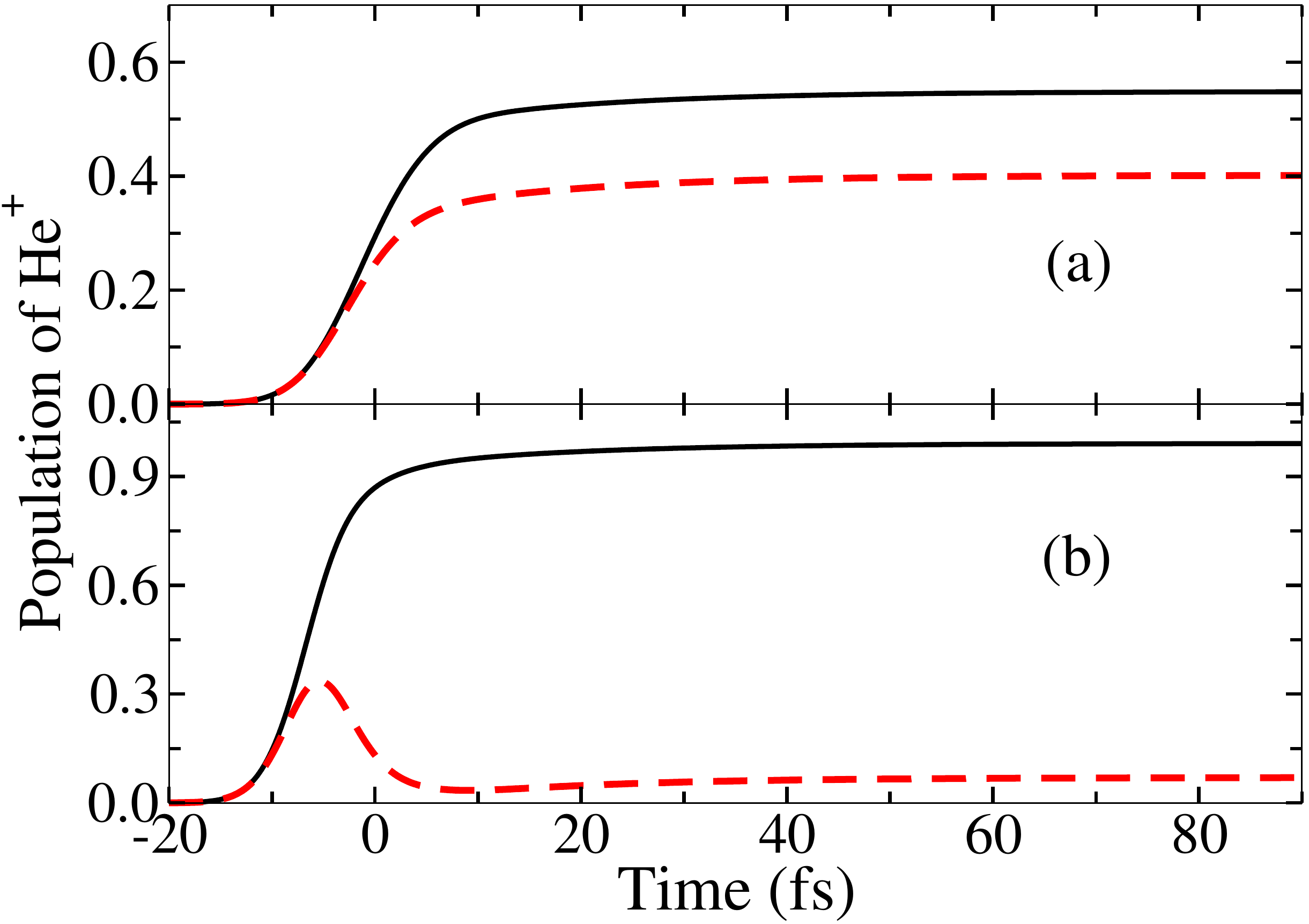}
\caption{Comparison of He$^+$ yield when its further ionisation is taken into account (red-dashed curves) and when it is not (black-solid curves), for the peak intensities $10^{14}$ W/cm$^2$-(plot (a)) and  $10^{15}$ W/cm$^2$-(plot (b)). The yields were obtained at the resonance frequency $60.154$ eV and for $\tau_p = 7$ fs, $\tau_c = 0.5$ fs ($M = 14$). 
}
\label{fig:avg_setA_time}
\end{figure}

In an actual experiment, further ionization of the singly-ionized He takes place and a proper treatment requires to take into account the complete set of the density matrix equations. The plots (d), (e) and (f) at the right side of the Fig.~(\ref{fig:avg_mc_setA}) include the ionisation channel from He$^+$ to He$^{+2+}$ ($\gamma_c = 0.436$~a.u. but $\gamma_a = 0$). The first thing to note is the broadened yield around resonance. In contrast, a deterministic (and Fourier-limited) pulse would have developed a sharp asymmetric peak near the resonance energy of 2s2p i.e., 60.154 eV (2.211 a.u.), typical of an AIS resonance with $q=-2.733$ Fano parameter. The width of this resonance would reflect the AIS width, $\Gamma_a$. Here, the fluctuations of the FEL field have smoothed this sharp resonance resulting to a broadened shape. 

On another note, in the same figure we can see that for lower peak intensities the He$^+$ yields obtained for $\gamma_c=0$ (left plots) are close to those when $\gamma_c=0.426$~a.u. (right plots). Considering only the right side plots of Fig.~(\ref{fig:avg_mc_setA}), we notice an increase in the yield value from (d) to (e), which follows the trend of (a) to (b). But the same trend is lost in (f). This is due to the enhanced He$^+$ generated at times before the pulse reached its peak, followed by its quick ionization to He$^{+2}$. Obviously this is not the case for lower peak intensities where He$^+$ is generated past the pulse's peak. This is shown clearly at the bottom panel of Fig.~(\ref{fig:avg_setA_time}) where as soon as the population of He$^+$ reaches a maximum, it decays rapidly into He$^{2+}$ for the highest peak intensity ($\sim 10^{15}$ W/cm$^2$); in contrast with what is observed at the upper panel of the same figure, where the He$^+$ populations develop post the pulse's peak value (assumed at $t=0$).

\subsection{Effects of the field's AC shape:}
We now investigate the effects of the shape of a Lorentzian-dependent AC coherence function on the lineshape and  compare with a square-exponential dependence. 

It is easily shown that a stationary (or nearly stationary) field of an exponentially-correlated coherence function has a Lorentzian-like power spectrum whereas for a square-exponentially dependent coherence function the spectrum is Gaussian. We want to see how the field's AC function affects the AIS lineshape; we compare the results for the two different AC shapes for the same coherence time, $\tau_c$ ; one exponentially correlated [see Eq.~(\ref{eq:c1-exponential})] and the other approximately close to the one produced from FEL sources, especially for pulses of longer duration \cite{Krinsky1,kim:2017}, being the square-exponential of Eq.~(\ref{eq:c1-gauss}).

\begin{figure}
 \includegraphics[width=65mm,scale=0.5, angle=270]{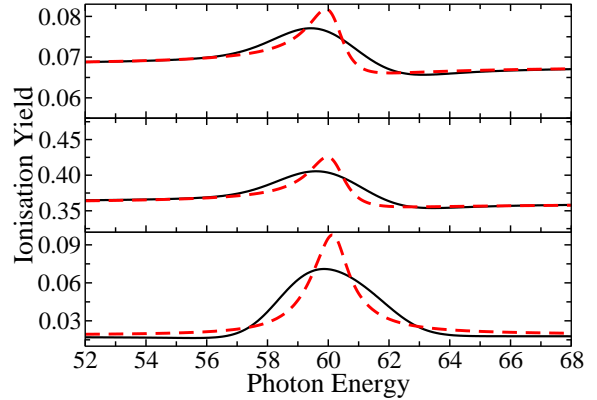}
\caption{Comparison of the ionisation yields obtained by using Gauss correlation (black solid curve) and Lorentz correlation (red dashed curve) for the peak intensities of $10^{13}$ W/cm$^2$, $10^{14}$ W/cm$^2$ and $10^{15}$ W/cm$^2$ from top to bottom, respectively. Other parameters are $\tau_p = 7$ fs, $\tau_c = 0.5$ fs ($M = 14$).}
\label{fig:gauss_lorentz}
\end{figure}

In Fig.~(\ref{fig:gauss_lorentz}) we plot the yields obtained for three different intensities, from top to bottom, the peak intensity increases as $10^{13}$ W/cm$^2$, $10^{14}$ W/cm$^2$ and $10^{15}$ W/cm$^2$ respectively. We notice that the peaks of yields have different values with the exponentially-correlated fields to result to a narrower shape. While the peaks at the resonance region differ, their tails almost match. Therefore, there is a notable difference in the yields in the resonance region which can be attributed exclusively in the choice of the AC temporal dependence.

\subsection{Effects of Field's correlation time on the AIS lineshape}

As we have already established the accuracy of the averaging technique, we may now consider a larger pulse duration of $\tau_p=45 $ fs (corresponding to $\tau_{FWHM}\approx 75$ fs) (corresponding to FWHM laser bandwidth of $0.4$ eV ) and study the effects of different correlation times of the field on the AIS lineshape using the averaging method. This pulse is close to parameters used in \cite{FEL_expt}.

\begin{figure}

\includegraphics[width=65mm,scale=0.5,angle=270]{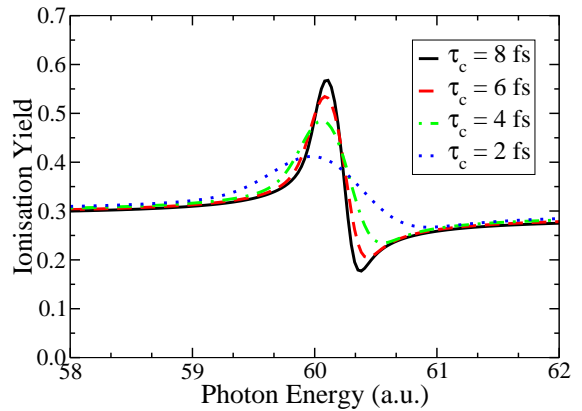}

\caption{Comparison of the ionisation yield obtained by using Gaussian form of the field's autocorrelation function at $10^{13}$ W/cm$^2$ for different coherence times $\tau_c$: 2 fs (black-curve), 4 fs (red-curve), 6 fs (green-curve) and 8 fs (blue-curve). At resonance, the yield is highest for the longest coherence time and it gradually decreases as the $\tau_c$ is decreased. The pulse duration used is $\tau_p = 45$ fs. }
\label{fig:effects_tau_c}
\end{figure}

In Fig.~(\ref{fig:effects_tau_c}), the ionisation yield obtained for various correlation times and a and peak intensity is $10^{13}$~W/cm$^2$. It is seen that, as the correlation time decreases the yield value at the resonance's position drops and the AIS lineshape becomes broader. $\tau_c$ is directly related to the laser bandwidth and therefore, a decrease in the correlation time of the field implies an increase in the laser bandwidth which manifests as the broadening of the AIS lineshape; so the broadening effect is compensated by the drop of the resonance peak. As, at the tails of the yield, the values are quite constant it is concluded that the coherence time effects are manifested mainly around the resonance peak. 
This behaviour is also consistent with the behaviour of the $S_{1t}$ in Eq.~(\ref{eq:s1ta}) where towards smaller $\tau_c$ the line-shape gradually broadens at the wings following a Lorentzian-like trend. 


\section{Conclusion}

We presented a cumulants-based perturbative method to derive the averaged EOMs of the mean density matrix elements which describe the resonant ionization of atomic systems under stochastic fields. We have applied its second-order truncated expansion for the near-resonant ionization (via He~$2s2p$ AIS state at $\sim 60.2$ eV) and compared with MC calculations. The convergence of the perturbation expansion depends on the field's strength and its coherence AC time; the basic assumption is that the coherence AC time is shorter than any other characteristic evolution time imposed by the field or the atomic structure itself (inverse Rabi-excitation amplitude and photoionization, autoionization width). We found that in general a Voight profile for the AIS is to be expected; this profile depends both on the field and the intensity AC functions. 

There are a number of other statistical aspects of the atom/FEL radiation interactions which necessitate further study in order to identify more comprehensively the role of the fluctuations in the ionization process. First, possible improvements of the averaging procedure needs to be investigated; one such alternative is based on the atomic physics projection-operator method, but now applied to stochastic problems, also leading to averaged equations for the density matrix elements, known as Nakajima-Zwanzig equation \cite{nakajima:1958,zwanzig:1960}; nevertheless this equation has a non-Markovian structure which implies knowledge of the average density matrix at all previous times, $\avop{\sigma(t^\prime)},t^\prime < t$ and not at one-time only, $\avop{\sigma(t)}$ as the Markov-type derived equation Eq.~(\ref{eq:tddm-averaged}).      

Along similar lines it is within our interests to improve the method and drop the assumption of $\tau_c$ as the shortest characteristic time. Another route we would like to focus is to provide a method where higher-order terms of the perturbative expansion can be calculated more automatically and efficiently for arbitrary FEL fields without the need to invoke the Gaussian statistics requirement; this will allow to model a larger class of problems than those that can be treated with the present method. Finally an alternative direction is to deal with problems where the statistics of the system's density matrix itself (beyond its average) is required; this calls for two-time averages of the type $\avop{\sigma(t) \sigma(t^\prime)}$ to be calculated; the method presented in this work, can be also applied for such averages to be calculated, of course within the limitations of the short-correlation time, $\tau_c$.
\section{Acknowledgements:}
T.K. wants to acknowledge the support by the Education, Audiovisual and Culture Executive Agency (EACEA) Erasmus Mundus Joint Doctorate Programme Project No. 2011 0033.
\appendix

\section{Evolution matrices}

If we define the near-resonant detuning, $\delta_0$, by
\begin{equation}
\delta_0 = E_a - E_g - \omega,
\label{eq:det_0}
\end{equation}
then the complex detunings $\hat{\Delta},\hat{\delta}$ are given as,
\begin{equation}
\hat{\delta} =  \imath \delta_0 + \frac{\Gamma_a}{2},
\qquad
\hat{\Delta} = \imath (s_a-s_g) + \frac{\gamma_g + \gamma_a}{2},
\label{eq:det_c}
\end{equation}
with $s_g,s_a$ the peak intensity ac-Stark shifts of the respective states. The interference matrix element is,
 \begin{equation}
\hat{\calD} =d_{ga}(1-\frac{i}{q_a}) = |D|e^{-\imath \phi_a}
 \label{eq:rabi}
\end{equation}
where $d_{ga}$ is the real part of the transition matrix element between the ground $|g\rangle$ and the excited $|a\rangle$ state and $q_a$ is the q-Fano parameter. We should emphasize the (within the Fano ionization picture) strict relation between the involved dynamics parameters: 
\begin{equation}
4|d_{ga}|^2 = q_a^2 \;\Gamma_a \gamma_g,
\label{eq:rabi-q-gamma}
\end{equation}
The constant matrices $\opL_i, i=0-2$ are as below,
\begin{equation}
\opL_0  = -
\begin{pmatrix}
0 & 0    & 0 & 0   \\ 
 0 & \Gamma_a &  0  & 0    \\ 
 0 & 0 & \hat{\delta} & 0 \\
 0 & 0 & 0  & \hat{\delta}^*
\end{pmatrix},\\
\label{eq:L0} 
\end{equation}

\begin{equation}
\opL_1  = \imath
\begin{pmatrix}
0  & 0    & - \hat{\calD} & 0   \\ 
 0 & 0 &  \hat{\calD}^*  & 0    \\ 
 0 & 0 & 0 & 0 \\
 \hat{\calD}^* & - \hat{\calD} & 0  & 0
\end{pmatrix}, 
\label{eq:L1}
\end{equation}
with $\opL_1^T$ representing the transpose of $\opL_1$. Finally,
\begin{equation}
\opL_2  = - 
\begin{pmatrix}
\gamma_g  & 0    & 0 & 0   \\ 
 0 & \gamma_a &  0  & 0    \\ 
 0 & 0 & \hat{\Delta} & 0 \\
 0 & 0 & 0  & \hat{\Delta}^*
\end{pmatrix}.
\label{eq:L2} 
\end{equation}
With the above definitions $\widetilde{L}_b(t)$ takes the following explicit form,

\begin{equation}
\widetilde{L}_b(t) =  
\begin{pmatrix}
-\gamma_g\widetilde{\calI}(t)  & 0    & - \imath \caltE(t)\hat{\calD} & \imath \caltE^*(t) \hat{\calD}^*                \\ 
         0 & -\gamma_a\widetilde{\calI}(t) &  \imath \caltE(t) \hat{D}^*  & -\imath \caltE^*(t) \hat{\calD}                              \\ 
 - \imath \caltE^*(t) \hat{\calD} & \imath \caltE^*(t) \hat{\calD}^* & \hat{\Delta} \widetilde{\calI}(t) & 0                 
\\
 \imath \caltE(t) \hat{\calD}^* & -\imath \caltE(t) \hat{\calD} & 0               & \hat{\Delta}^{^*} \widetilde{\calI}(t) 
\end{pmatrix}. 
\end{equation}

\section{Partially-ordered cumulants}

In this appendix we calculate the first four expansion terms of the averaged equation Eq.~(\ref{eq:tddm-averaged}) by explicit calculation of the $\dot{K}(t)$ operator in Eq.~(\ref{eq:dK_dt}). Let's consider the average evolution operator, $\widetilde{U}(t)$ of Eq.~(\ref{eq:dyson_v}):
\begin{equation}
\avop{\widetilde{U}(t)} = \opOne + \sum_{n=1} U_n(t) 
\label{eq:dyson_moments_v0}
\end{equation}
with the $U_n(t)$ terms defined in Eq.~(\ref{eq:moments-1}). The time-derivative of $U_n(t)$ is given by: 
\begin{equation}
\dot{U}_n(t) = \int_{t_i}^t dt_1 \int_{t_i}^{t_1} dt_2 ..\int^{t_{n-2}}_{t_i} \!\!\!dt_{n-1} M_n(t,t_1,\cdots t_{n-1}).  
\label{eq:dyson_moments_v1}
\end{equation}
with $M_n(t_1,\cdots, t_n)$ given in Eq.~(\ref{eq:moments}). The integral limits ensure the prescribed chronological order $t > t_{1} > t_2 > \cdots > t_n>t_i$. For non-commutative matrices, the chosen evaluation order in the integral matters and was taken into consideration by introducing the  proper chronological (Dyson) operator in the time derivative of $K(t)$:
\begin{align*}
\dot{K}(t) &= \hat{T}\left[\frac{\avop{\dot{\widetilde{U}}(t)}}{\avop{\widetilde{U}(t)}}\right] 
= \hat{T}\left[ \frac{ \sum_n \dot{U}_n}{\opOne + \sum_n U_n} \right] 
\\
&=\hat{T} \left[ 
\left(\sum_n \dot{U}_n \right) \left( \opOne - \sum_n U_n + (\sum_n U_n) ^2 - \cdots \right)
\right]. 
\end{align*}
In the second line of the last equation a Neumann expansion of the denominator was performed. This is a legitimate expansion, provided that $|\sum_n U_n| < 1$; an assumption which should hold if the elements $U_n$ are to be used to make up an evolution operator. Expanding the terms of the last line inside the $\hat{T}$ bracket and gathering together those of the same order we have (up to 4th power are shown below) : 
\begin{align*}
& (\dot{U}_1 + \dot{U}_2 + \cdots)(\opOne - U_1 - U_2 - \cdots + U_1^2 + U_2^2 + \cdots )
\\
&=
\dot{U}_1 + \dot{U}_2 + \dot{U}_3 + \dot{U}_4 
- \dot{U}_1 U_1 - \dot{U}_1 U_2 -\dot{U}_1 U_3 
\\
&- \dot{U}_2 U_1 - \dot{U}_2 U_2 \cdots - \dot{U}_3 U_1 - \cdots
\\
&+ \dot{U}_1 U_1^2 + \dot{U}_1 U_1 U_2 +  \dot{U}_1 U_2 U_1 + \cdots
\\
&=  \dot{U}_1 + \left( \dot{U}_2 - \dot{U}_1 U_1 \right) 
+
\left( \dot{U}_3 -  \dot{U}_1 U_2 -\dot{U}_2 U_1  +  \dot{U}_1 U_1^2 \right)
\\
&+
\left( \dot{U}_4 - \dot{U}_2 U_2 - \dot{U}_1 U_3 - \dot{U}_3 U_1 
+ \dot{U}_1 U_1 U_2 +  \dot{U}_1 U_2 U_1 
\right)  
\\
&+ \cdots + O(U^5)
\end{align*}
So we may write compactly,$ \dot{K}(t) = \sum_{n=1}^{\infty} \dot{\kappa}_n(t)$
with,
\begin{align}
\dot{\kappa}_1(t) & = \hat{T}\left[ \dot{U}_1 \right]  = 0 
\label{eq:k1dot}
\\
\dot{\kappa}_2(t) & = \hat{T} \left[\dot{U}_2 - \dot{U}_1U_1  \right] 
= \hat{T}\left[\dot{U}_2\right] \nonumber \\
& = \int_{t_i}^{t} dt_1 \,M_2(t,t_1) 
\label{eq:k2dot}
\\
\dot{\kappa}_3(t) & = \hat{T} \left[ \dot{U}_3 -  \dot{U}_1 U_2 -\dot{U}_2 U_1  +  \dot{U}_1 U_1^2 \right]
=  \hat{T} \left[ \dot{U}_3\right]  
\nonumber \\
&= \int_{t_i}^{t}dt_1 \int_{t_i}^{t_1} dt_2 \; M_3(t,t_1,t_2), 
\label{eq:k3dot}
\end{align}
where due to our special choice of $\avop{\widetilde{L}(t)} = 0$ we set $U_1= \dot{U_1}=0$. In addition the chronological operator becomes redudant since the integral limits take care automatically the chosen prescription. For the $\dot{k}_4(t)$ term some further algebra is required though:
\begin{align}
\dot{\kappa}_4(t) & = 
 \hat{T} \left[  \dot{U}_4 - \dot{U}_2 U_2 \right]  
 \label{eq:k4dot-tmp}
 \\
 &=
\hat{T}\left[ 
\int_{t_i}^{t} dt_1 \int_{t_i}^{t_1} dt_2 \int_{t_i}^{t_2} dt_3 \;M_4(t,t_1,t_2,t_3)
\right]
\nonumber
\\
& - 
\hat{T}
\left[
\int_{t_i}^{t} dt_1  M_2(t,t_1) \int_{t_i}^{t} dt_2  \int_{t_i}^{t_2} dt_3\; M_2(t_2,t_3)
\right]
\nonumber
\\
&= \int_{t_i}^{t} dt_1 \int_{t_i}^{t_1} dt_2 \int_{t_i}^{t_2} dt_3 \;M_4(t,t_1,t_2,t_3) 
\nonumber
\\
&
-
\hat{T}
\left[
\int_{t_i}^{t} dt_1 \int_{t_i}^{t} dt_2 \int_{t_i}^{t_2} dt_3\; M_2(t,t_1)   M_2(t_2,t_3)
\right]
\nonumber
\end{align}
It is not straightforward to drop the $\hat{T}$ operator from the last line since the integral limits do not coincide with the prescribed time-ordering $t>t_1>t_2>t_3>t_i$. Some manipulations are needed to arrive at a triple-integral with the proper limits. To this end we make use of the following identity:
\begin{equation}
\int_b^a dx \int_x^a dy f(x,y) = \int_b^a dx \int_b^x dy f(y,x). 
\label{eq:double_integral}
\end{equation}
Adopting temporarily the notation, $M_2(t_i,t_j) = M_{ij}$ and droping the $\hat{T}$ operator each time that conforms with the integral limits, we have for the last line of Eq.~(\ref{eq:k4dot-tmp}):
\begin{widetext}
\begin{align*}
&
\hat{T}
\left[
\int_{t_i}^{t} dt_1 \int_{t_i}^{t_1} dt_2 \int_{t_i}^{t_2} dt_3\; M_{01}   M_{23}
\right]
+
\hat{T}
\left[
\int_{t_i}^{t} dt_1 \int_{t_1}^{t} dt_2 \int_{t_i}^{t_2} dt_3\; M_{01}M_{23}
\right]
\\
&=
\int_{t_i}^{t} dt_1 \int_{t_i}^{t_1} dt_2 \int_{t_i}^{t_2} dt_3\; M_{01}M_{23}
+
\hat{T}
\left[
\int_{t_i}^{t} dt_1 \int_{t_1}^{t} dt_2 \int_{t_i}^{t_1} dt_3\; M_{01}M_{23}
\right] 
+
\hat{T}
\left[
\int_{t_i}^{t} dt_1 \int_{t_1}^{t} dt_2 \int_{t_1}^{t_2} dt_3\; M_{01}  M_{23}
\right]
\\
&=
\int_{t_i}^{t} dt_1 \int_{t_i}^{t_1} dt_2 \int_{t_i}^{t_2} dt_3\; M_{01}M_{23}
+
\int_{t_i}^{t} dt_1 \int^{t_1}_{t_i} dt_2 \int_{t_i}^{t_2} dt_3\; M_{02}M_{13} 
+
\hat{T}
\left[
\int_{t_i}^{t} dt_1 \int^{t_1}_{t_i} dt_2 \int_{t_2}^{t_1} dt_3\; M_{02}  M_{13}
\right]
\\
&=
\int_{t_i}^{t} dt_1 \int_{t_i}^{t_1} dt_2 \int_{t_i}^{t_2} dt_3\; M_{01}M_{23}
+
\int_{t_i}^{t} dt_1 \int^{t_1}_{t_i} dt_2 \int_{t_i}^{t_2} dt_3\; M_{01}M_{23} 
+
\int_{t_i}^{t} dt_1 \int_{t_i}^{t_1} dt_2 \int_{t_1}^{t_2} dt_3\; M_{03}  M_{12},
\end{align*}
\end{widetext}
 Now we can replace the integral of Eq.~(\ref{eq:k4dot-tmp}) and arrive at the final expression:
\begin{widetext}
\begin{equation}
\dot{\kappa}_4(t) = 
\int_{t_i}^{t} dt_1 \int_{t_i}^{t_1} dt_2 \int_{t_i}^{t_2} dt_3
\left[
M_4(t,t_1,t_2,t_3) - M_2(t,t_1)M_2(t_2,t_3) - M_2(t,t_2)M_2(t_1,t_3) - M_2(t,t_3)M_2(t_1,t_2)
\right]
\label{eq:k4dot}
\end{equation}
\end{widetext}
Note that for stochastic fields with Gaussian statistics and in the case of commutative matrices, we would have $\dot{\kappa}_3 = \dot{\kappa}_4 = 0 $ if we take into account the properties of the non-vanishing multitime moments \cite{vankampen:1992}:
\begin{equation}
M_{ijkl\cdots mn} = \sum M_{i'j'}M_{k'l'} \cdots M_{m'n'}, \quad n\quad \text{even}
\label{eq:gaussian_moments}
\end{equation}
where $(i',j'),(k',l'), \cdots (m'n')$ pairs are all possible combinations of the indices $(i,j,k,l,\cdots m,n)$.  

One may continue and work out the higher-order cumulants expressions along similar lines if they are needed. Nevertheless, alternative recipes (not less laborious, though) for their calculations in terms of the moments may also be found be found in the original literature \cite{kubo:1962,vankampen:1992}.

\bibliographystyle{apsrev4-2}
\bibliography{Thesis}

\end{document}